\documentclass[twocolumn,superscriptaddress,
amsmath,amssymb,aps,pra]{revtex4-1}

\usepackage{graphicx}
\usepackage{physics}
\usepackage{color}
\usepackage{dcolumn}
\usepackage{bm} 
\usepackage{float}
\usepackage{xcolor}
\usepackage{multirow}
\usepackage{tabularx}
\usepackage{comment}
\usepackage{mathtools}
\usepackage{booktabs}
\usepackage{amsthm}

\usepackage[colorlinks,urlcolor=blue,citecolor=blue,linkcolor=blue]{hyperref}

\newcommand{\be}{\begin{equation}}
\newcommand{\ee}{\end{equation}}
\newcommand{\vp}{{\vectorbold{p}}}

\newcommand{\vpd}{\vectorbold{p'}}
\newcommand{\vq}{\vectorbold{q}}
\newcommand{\vQ}{\vectorbold{Q}}
\newcommand{\vk}{\vectorbold{k}}
\newcommand{\vkd}{\vectorbold{k}'}
\newcommand{\eh}{\hat{e}^{}}
\newcommand{\ehd}{\hat{e}^{\dagger}}
\newcommand{\nn}{\nonumber}

\newcommand{\up}{\uparrow}
\newcommand{\down}{\downarrow}
\renewcommand{\k}{{\bf k}}
\newcommand{\p}{{\bf p}}

\newcommand{\sch}{Schr{\"o}dinger }

\begin{document}

\title{Efficient calculation of trion energies in monolayer transition metal dichalcogenides}

\author{Sangeet S. Kumar}
\affiliation{School of Physics and Astronomy, Monash University, Victoria 3800, Australia}
\affiliation{ARC Centre of Excellence in Future Low-Energy Electronics Technologies, Monash University, Victoria 3800, Australia}

\author{Brendan C. Mulkerin}
\affiliation{School of Physics and Astronomy, Monash University, Victoria 3800, Australia}
\affiliation{ARC Centre of Excellence in Future Low-Energy Electronics Technologies, Monash University, Victoria 3800, Australia}

\author{Antonio Tiene}
\affiliation{School of Physics and Astronomy, Monash University, Victoria 3800, Australia}
\affiliation{Departamento de F\'isica Te\'orica de la Materia
  Condensada, Universidad
  Aut\'onoma de Madrid, Madrid 28049, Spain}
\affiliation{Condensed Matter Physics Center (IFIMAC), Universidad Autónoma de Madrid, 28049 Madrid, Spain}

\author{Francesca Maria Marchetti}
\affiliation{Departamento de F\'isica Te\'orica de la Materia
  Condensada, Universidad
  Aut\'onoma de Madrid, Madrid 28049, Spain}
\affiliation{Condensed Matter Physics Center (IFIMAC), Universidad Autónoma de Madrid, 28049 Madrid, Spain}

\author{Meera M. Parish}
\affiliation{School of Physics and Astronomy, Monash University, Victoria 3800, Australia}
\affiliation{ARC Centre of Excellence in Future Low-Energy Electronics Technologies, Monash University, Victoria 3800, Australia}

\author{Jesper Levinsen}
\affiliation{School of Physics and Astronomy, Monash University, Victoria 3800, Australia}
\affiliation{ARC Centre of Excellence in Future Low-Energy Electronics Technologies, Monash University, Victoria 3800, Australia}

\date{\today}

\begin{abstract}
The reduced dielectric screening in atomically thin semiconductors leads to remarkably strong electron interactions. As a result, bound electron-hole pairs (excitons) and charged excitons (trions), which have binding energies in the hundreds and tens of meV, respectively, typically dominate the optical properties of these materials. However, the long-range nature of the interactions between charges represents a significant challenge to the exact calculation of binding energies of complexes larger than the exciton. Here, we demonstrate that the trion binding energy can be efficiently calculated directly from the three-body Schr\"odinger equation in momentum space. Key to this result is a highly accurate way of treating the pole of the electronic interactions at small momentum exchange (i.e., large separation between charges) via the Land\'e subtraction method. Our results are in excellent agreement with quantum Monte Carlo calculations, while yielding a substantially larger ratio of the trion to exciton binding energies than obtained in recent variational calculations. Our numerical approach may be extended to a host of different few-body problems in 2D semiconductors, and even potentially to the description of exciton polarons.
\end{abstract}

\maketitle
\section{Introduction}
 
Two-dimensional (2D) materials have gained significant attention due to their unique electronic, optical, and mechanical properties, and their potential for applications in next-generation technologies such as nanoelectronics, photonics, sensing, energy storage, and optoelectronics~\cite{Britnell2013,Schaibley2016}. Many of these properties can be attributed to the presence of strongly bound few-body complexes, such as excitons (electron-hole bound states) and trions (bound states of an exciton with an electron or a hole), which arise from the reduced screening of the Coulomb interaction due to quantum confinement and the relatively heavy carrier masses in these materials~\cite{Velicky2017}. In monolayer transition-metal dichalcogenides (TMDs), exciton binding energies have been found to be several hundred meV~\cite{He2014,Chernikov2014}, while trion binding energies have been reported to reach values up to $20$ to $40$~meV~\cite{Mak2013,Ross2013,Wang-Urbaszek_PRB2014,BZhu2015,Yang2015,Plechinger_PSS2015,Singh2016,Arora2019} (see Refs.~\cite{WangRev2018,Durnev2018} for recent reviews). These values are substantially larger than in traditional quantum wells, such as GaAs~\cite{Finkelstein1996,Bracker_PRB2005,Joseph2005} and CdTe~\cite{Kheng1993,Huard_PRL2000,Portella2004,Moody2014}, where the trion binding energy is of the order of $0.5-3$~meV, and ZnSe, where it can reach $10$~meV~\cite{Astakhov2002}. Larger binding energies are desirable as they prevent thermal dissociation, thus increasing their stability at room temperature, which is important for practical applications~\cite{Mueller2018,Emmanuele2020}. 
  
From a theoretical perspective, calculations of the trion bound state are challenging due to the singular nature of the electronic interaction, both at long and short distances. For the case of unscreened Coulomb interactions in three dimensions, a precise treatment of the long-range divergence in momentum space was proposed by A.~Land{\'e} and realized by Kwon and Tabakin in their work on hadronic atoms~\cite{Kwon1978}. The key idea is to subtract and add an appropriately chosen factor to the interaction potential in order to produce a well-behaved effective potential in the numerical solution. This method has been tried and tested in the treatment of the Coulomb two-body problem both in the context of nuclear physics~\cite{Landau1983,Norbury1994,Ivanov2001} and 2D semiconductors~\cite{Laird2022}, and is numerically inexpensive and accurate. The technique can also be applied to the Rytova-Keldysh potential~\cite{deLaFuentePico2025}, which accounts for the spatially inhomogeneous dielectric screening of Coulomb interactions in a monolayer TMD~\cite{Rytova1967,Keldysh1979,Cudazzo2011}. 

In this paper, we apply the Land{\'e} subtraction technique to the calculation of trion binding energies in two-dimensional semiconductors, focusing on parameters relevant to monolayer TMDs as well as traditional quantum wells. We supplement the subtraction scheme with a simple iterative Lanczos-type method~\cite{Stadler1991,Demmel1997,Hadizadeh2007,Hadizadeh2012}, which allows us to calculate the trion energy directly from the momentum-space \sch equation. For the case of equal electron and hole masses and uniform dielectric screening---which, due to its high symmetry, is the most investigated scenario---we find the trion binding energy to be $0.122 \varepsilon_X$ (with $\varepsilon_X$ the exciton binding energy). This indicates a stronger binding than that obtained with variational wave functions which predict trion binding energies $\sim 0.11 \varepsilon_X$~\cite{Sergeev2001,Courtade2017}, and agrees well with  quantum Monte Carlo (QMC) calculations which predict $\sim 0.12\varepsilon_X$~\cite{Szyniszewski2017}. More generally, we calculate the trion binding energy in monolayer TMDs as a function of the dielectric screening length~\cite{Rytova1967,Keldysh1979}, again finding excellent agreement with QMC calculations~\cite{Szyniszewski2017} while obtaining a larger trion binding energy than in variational calculations~\cite{Courtade2017}. We also find very good agreement with the exact diagonalization calculations of Ref.~\cite{Fey2020} for the select TMD parameters chosen in that work. Due to their general nature, our numerical methods can be applied to a host of other few- and many-body problems in 2D semiconductors.

The paper is organized as follows. The model is introduced in Sec.~\ref{Model} and we demonstrate the Land{\'e} subtraction technique in Sec.~\ref{sec:Lande} in the context of the exciton problem. We introduce the trion Schr\"odinger equation in Sec.~\ref{TrionSect}, where we also apply the Land{\'e} subtraction technique and solve for the trion binding energies using a Lanczos-type algorithm. We conclude and give a brief outlook in Sec.~\ref{sec:conc}.

\section{Model} \label{Model}
To describe the behavior of a few electrons and holes in a 2D semiconductor, we use the following Hamiltonian
\begin{align}\label{eq:Ham}
\hat{H} 
 = &  \underset{\vk\sigma}{\sum} \left( 
 \epsilon^{e}_{\vk} \, \ehd_{\sigma,\vk} \eh_{\sigma,\vk} +
 \epsilon^{h}_{\vk} \, \hat{h}^{\dagger}_{\sigma,\vk} \hat{h}_{\sigma,\vk}^{} \right) 
\nonumber \\ 
&
\hspace{.2cm} +\frac{1}{2} \underset{\vk \vkd \vq \sigma \sigma'}{\sum} V_{\vq} \left(
\ehd_{\sigma,\vk+\vq} \ehd_{\sigma',\vkd-\vq} \eh_{\sigma',\vkd} \eh_{\sigma,\vk} 
\nonumber \right.\\
&
\hspace{2.5cm}
\left. +\hat{h}^{\dagger}_{\sigma,\vk+\vq} \hat{h}^{\dagger}_{\sigma',\vkd-\vq} \hat{h}_{\sigma',\vkd}^{} \hat{h}_{\sigma,\vk}^{}  \nonumber \right. \\
& 
 \left. \hspace{2.5cm}-2 \ehd_{\sigma,\vk+\vq}\hat{h}^{\dagger}_{\sigma',\vkd-\vq}\hat{h}_{\sigma',\vkd}^{} \eh_{\sigma,\vk}
\right) .
\end{align}
Here, and in the following, we work in units where the system area and $\hbar$ are both set to unity, and we take the continuum limit of sums as $\sum_{\vk}\to\int \frac{d\vk}{(2\pi)^2}$. In Eq.~\eqref{eq:Ham}, the creation (annihilation) operators of spin-$\sigma$ electrons and holes at momentum $\vk$ are written as $\ehd_{\sigma,\vk} \text{ } (\hat{e}_{\sigma,\vk}^{})$ and $\hat{h}^\dagger_{\sigma,\vk} \text{ } (\hat{h}_{\sigma,\vk}^{})$, respectively, and similarly, the kinetic energies of the electrons and holes as $\epsilon^{e}_{\vk}=\frac{|\vk|^2}{2m_{e}}\equiv \frac{k^2}{2m_{e}}$ and $\epsilon^{h}_{\vk}=\frac{k^2}{2m_{h}}$ (throughout this work we measure the energies with respect to the bandgap energy). For simplicity, we use an effective mass approximation for the description of electrons and holes. This is motivated by comparisons between variational~\cite{Berkelbach2013} and first principles calculations~\cite{Ramasubramaniam_PRB2012,Komsa_PRB2012,Shi_PRB2013} that have shown this approximation to work well for  excitons in monolayer TMDs~\cite{Berkelbach2013}. Similarly, there is a growing body of literature showing remarkable agreement between effective-mass models for excitons in these systems and experiment --- see, e.g., Ref.~\cite{Zipfel-Chernikov-MF_PRB2018}. We stress that the effective mass approximation is expected to improve in accuracy for weaker bound states, such as trions, since their wave functions are dominated by low momenta at the bottom of the energy bands.

The interactions between charged particles in a 2D semiconductor are described by Coulomb interactions. In the case of spatially uniform dielectric screening of the interactions, the potential takes the usual form 
\begin{equation}
\label{eq:Vcoulomb}
V^{C}_{\vk} = 
\frac{e^2}{4\pi\epsilon}\frac{2\pi}{k}, 
\end{equation}
in momentum space. Here, $e$ is the electron charge and $\epsilon\equiv\varepsilon\epsilon_0$ the permittivity of the material, defined in terms of the dielectric constant $\varepsilon$ and the vacuum permittivity $\epsilon_0$. This potential is typically an excellent approximation in conventional quantum wells, such as III-V and II-VI heterostructures. In atomically thin materials, the spatially inhomogeneous dielectric screening of the monolayer instead results in the Rytova-Keldysh potential~\cite{Rytova1967,Keldysh1979}
\begin{equation} \label{eq:VRK}
V^{RK}_{\vk} =  \frac{e^2}{4\pi\epsilon} \frac{2\pi}{k(1+r_0 k)}, 
\end{equation}
where $r_0$ is the dielectric screening length and the permittivity $\epsilon$ is now the average of those of the surrounding regions above and below the monolayer: $\epsilon\equiv \varepsilon\epsilon_0=(\varepsilon_1+\varepsilon_2)\epsilon_0/2$. For a TMD monolayer of thickness $d$ and dielectric constant $\varepsilon_{ML}$, the screening length is defined as $r_0=d \varepsilon_{ML}/(2\varepsilon) $~\cite{Berkelbach2013}.

In position space, the two potentials take the forms:
\begin{subequations}
\begin{align}
    V^C(r) &= \frac{e^2}{4\pi\epsilon}\frac1r
    ,\\
    V^{RK} (r) &= \frac{e^2}{4\pi\epsilon}\frac{\pi}{2r_0}
    \left[H_0\left(\frac{r}{r_0}\right)-Y_0\left(\frac{r}{r_0}\right)\right],
\end{align}
\end{subequations}
where $H_0(x)$ is the zeroth-order Struve function and $Y_0(x)$ the zeroth-order Bessel function of the second kind. The Rytova-Keldysh potential has the same behavior at large separation $r$ between charges (small momentum exchange) as the 2D uniform Coulomb potential~\eqref{eq:Vcoulomb}. However, the potential is strongly modified at small separation between the charges,  corresponding to large momentum exchange: While the usual Coulomb potential scales as $-1/r$ in this limit, the Rytova-Keldysh potential scales as $\ln r$. In both cases, the long-range nature of the potential leads to the divergence seen in Eqs.~\eqref{eq:Vcoulomb} and \eqref{eq:VRK} when $k \rightarrow 0$, which is challenging to handle precisely in numerical calculations. As we explain in Section~\ref{sec:Lande}, this pole can be efficiently treated using a scheme adopted from nuclear physics~\cite{Kwon1978}.

In the following, we will relate the strength of the Coulomb potential to the reduced electron-hole mass $m_r=m_em_h/(m_e+m_h)$ and the effective 2D Bohr radius $a_0$ via $\frac{1}{2m_{r} a_{0}} \equiv \frac{e^2}{4\pi\epsilon}$). This choice of system parameters ensures that all of our results for the binding energies of excitons and trions are universal functions of just two dimensionless ratios, $r_0/a_0$ and $m_e/m_h$. It is useful to also introduce the 2D Rydberg constant, $R_X\equiv\frac1{2m_ra_0^2}$, which corresponds to the $1s$ exciton binding energy at $r_0=0$, and the $1s$ exciton binding energy $\varepsilon_X\equiv |E_{1s}|$, at any value of $r_0$.

\section{Exciton and subtraction scheme} \label{sec:Lande}

\begin{figure}
\begin{center}
\includegraphics[width=0.60\linewidth]{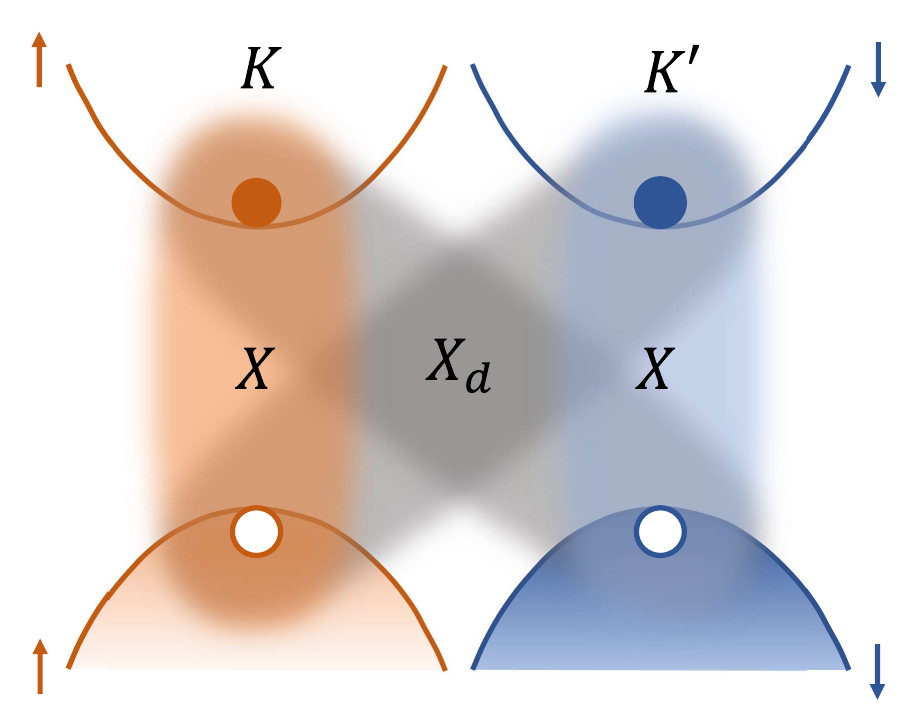}
\caption[system]{\label{fig:ExcitonFig1} Schematic illustration of direct ($X$) and dark ($X_d$) excitons in monolayer TMDs, such as MoSe$_2$. We explicitly show with shaded regions the $K$ (orange) and $K'$ (blue) valley direct excitons, as well as the dark excitons as gray shaded regions. Since our model does not include electron-hole exchange, these exciton states are degenerate. 
}
\end{center}
\end{figure}

\begin{figure}[t]
\begin{center}
\includegraphics[width=1.00\linewidth]{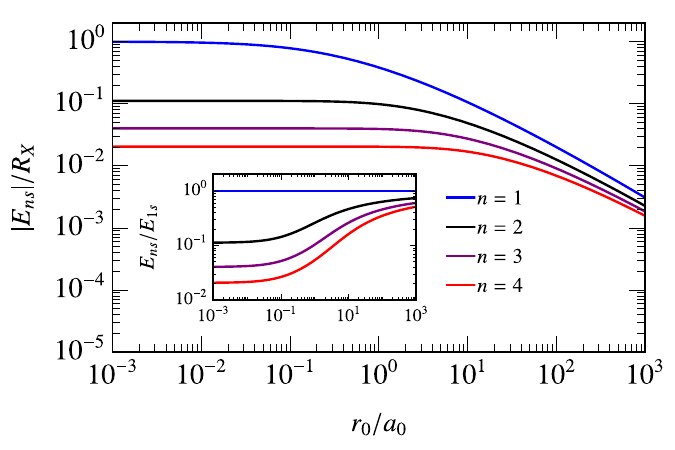}
\caption[system]{\label{fig:Screenedfig:ExcitonFig} 
Binding energy of the first few $ns$ exciton states as a function of the screening length $r_0$ in the Rytova-Keldysh potential. The inset shows the ratio of the energies of the excited excitons to that of the ground state. While $r_0/a_0$ is typically negligibly small in traditional quantum wells, it is commonly between 10 and 100 in atomically thin semiconductors.
}
\end{center}
\end{figure}

We now illustrate the Land{\'e} subtraction technique by applying it to the exciton Schr\"{o}dinger equation~\cite{Laird2022}. Defining a given exciton state as $\ket{X} = \sum_{\vk} \psi_{\vk} \ehd_{\vk,\sigma} \hat{h}^{\dagger}_{-\vk,\sigma'} \ket{0}$, the \sch equation $\hat H\ket{X}=E\ket{X}$ takes the form
\begin{equation} \label{eq:ExcitonSE}
(E-\epsilon^e_{\vp}-\epsilon^h_{\vp}) \psi_{\vp} = - \underset{\vk}{\sum} V_{\vp-\vk} \psi_{\vk} .
\end{equation}
Note that since we do not consider electron-hole exchange, the wave function $\psi _{\vk}$ is independent of spin. As such, Eq.~\eqref{eq:ExcitonSE} describes the pairing of an electron and a hole, where either of them can have spin $\up$ or $\down$. We schematically illustrate both the optically active direct exciton and the optically dark states in, e.g., a MoSe$_2$ monolayer, in 
Fig.~\ref{fig:ExcitonFig1}.

The essence of Land{\'e}'s scheme is to subtract a quantity which cancels out the simple pole of the Coulomb potential numerically, while adding the same quantity (semi) analytically~\cite{Kwon1978,Landau1983,Chuang1991,Norbury1994,Ivanov2001}. We therefore subtract and add the quantity $\psi_{\vp} \sum_{\vk} g_{\vp,\vk}$ from Eq.~\eqref{eq:ExcitonSE} to obtain 
\begin{align} \label{eq:ExcitonSESubtracted}
\left(E-\epsilon^e_{\vp}-\epsilon^h_{\vp}\right) \psi_{\vp} = - \sum_{\vk} \left(V_{\vp-\vk} \psi_{\vk} -g_{\vp,\vk} \psi_{\vp}\right)
-  K_{p} \psi_{\vp},
\end{align}
with $ K_{p} \equiv \sum_{\vk} g_{\vp,\vk}$. Here, $g_{\p,\k}$ should be chosen such that it exactly cancels the pole at $\k=\p$ when the term in parentheses on the right hand side of Eq.~\eqref{eq:ExcitonSESubtracted} is evaluated numerically. Obviously, there is significant additional freedom to choose $g_{\p,\k}$ such that it aids the convergence at large momenta, and ideally such that $K_p$ can be evaluated analytically. A convenient choice, identified in Ref.~\cite{Kwon1978} and applied to the exciton problem in Refs.~\cite{Chuang1991,Laird2022,deLaFuentePico2025}, is to take  $g_{\vp,\vk}=\frac{2p^2}{p^2+k^2} V_{\vp-\vk}$. In this case, crucially, $K_{p}$ can be obtained analytically for the Coulomb interaction ($V^{C}$), and partially analytically for the Rytova-Keldysh interaction ($V^{RK}$). This yields, respectively,
\begin{subequations}
\begin{align}
K^{C}_p &=\frac{p}{2m_r a_0}  \frac{1}{2\sqrt{2}}\frac{\Gamma(1/4)^2}{\sqrt{\pi}}, \\
K^{RK}_p
&=K^{C}_p -\frac{p}{2m_r a_0}   
\int^{\infty}_{0} 
\frac{2p r_0 t}{1 + \sqrt{2}p r_0 t }
\frac{dt}{\sqrt{1+t^4}},
\end{align}
\end{subequations}
where $\Gamma(x)$ is the Euler gamma function. This process removes the singularity, producing a well-behaved effective potential within the numerical integration, and thus drastically speeding up convergence. We can now use a Gaussian quadrature grid to convert Eq.~\eqref{eq:ExcitonSESubtracted} to an eigenvalue matrix equation, from which the energies and wave functions of the excitonic $s$-wave series can be readily calculated~\cite{NumericRec}. 

In Fig.~\ref{fig:Screenedfig:ExcitonFig}, we show the energy of the first few $s$-wave Rydberg exciton states  computed using the Rytova-Keldysh potential~\eqref{eq:VRK} as a function of the screening length. At $r_0=0$ we recover precisely the usual 2D Rydberg series of excitons, with energies $E_{ns}=-\frac{R_X}{(2n-1)^2}$. As $r_0$ increases, we see that in units of the Rydberg constant $R_X$ the presence of $r_0$ decreases the exciton binding energy. However, we note that in absolute units, excitons are weaker bound in quantum well structures than in atomically thin semiconductors due to the stronger overall screening. Furthermore, the deepest bound states deviate first from the hydrogen-like result, which is due to how the screening length $r_0$ only modifies the Coulomb potential at short electron-hole separation. Our results agree well with previous calculations for excitons in monolayer TMDs, such as Ref.~\cite{Berkelbach2013}. We also remark that our numerics converge very rapidly, with all the data points for the ground state ($n=1$, blue line) in Fig.~\ref{fig:Screenedfig:ExcitonFig} requiring only 16 momentum and 8 angular grid points, while the third excited state ($n=4$, red line) converges similarly with 52 and 26 momentum and angular grid points.

\section{Trion} \label{TrionSect}

\begin{figure*}[t]
\includegraphics[width=.8\linewidth]{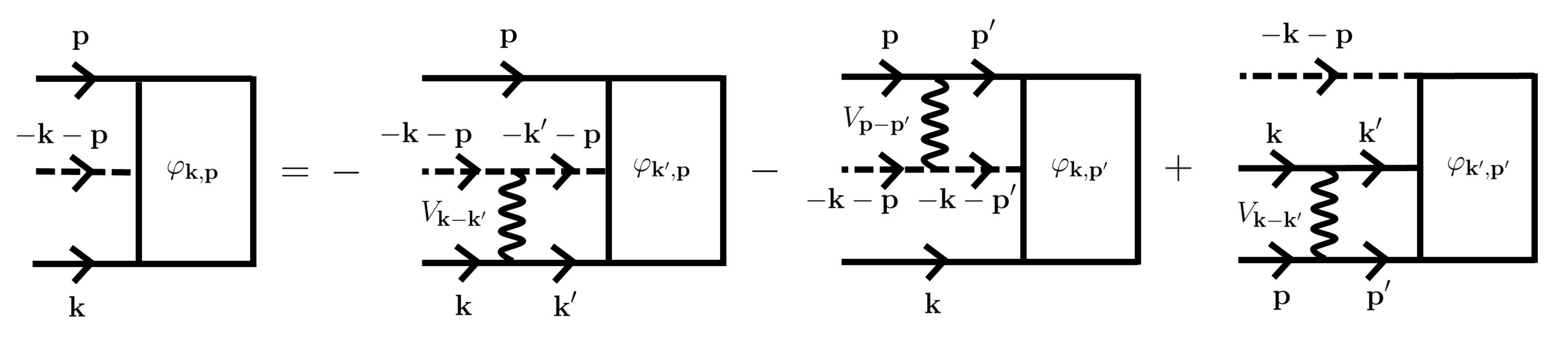}
\caption[system]{\label{TrionInteractionFig}
Diagrammatic representation of the trion Schr\"{o}dinger equation Eq.~\eqref{eq:TrionSEFull}, with the two distinguishable electrons drawn as solid lines, the hole as a dashed line, and the potential between the interacting particles as a wavy line. The first and second terms on the right-hand side represent the interaction between the hole and one of the two electrons, and the third term the electron-electron interaction.
We have used the transformation $\varphi_{\vk,\vp} =  \chi_{\vk,\vp}/(E-\epsilon^{e}_{\vk}-\epsilon^{e}_{\vp}-\epsilon^{h}_{\vk+\vp})$ to aid the correspondence to Eq.~\eqref{eq:TrionSEFull}.
}
\end{figure*}

We now turn to the main topic of this work, the calculation of trion binding energies in 2D semiconductors. Motivated by the advent of monolayer TMDs which feature very large trion binding energies, this has recently been studied in a number of works with a variety of numerical techniques~\cite{Kezerashvili_FBSReview2019}. These include variational methods~\cite{Berkelbach2013,Courtade2017} (see also older works on quantum well systems~\cite{Usukura1999,Sergeev2001,Sergeev_Nanotechnology2001}), mapping the three-body problem in two dimensions onto one particle in a three-dimensional potential~\cite{Ganchev2015}, the stochastic variational method~\cite{Kidd_PRB2016}, diffusion QMC~\cite{Mayers2015,Spink2016,Szyniszewski2017}, ab initio calculations~\cite{Deilmann2017}, finite element methods with multiband and effective mass approximations \cite{Donck2017}, exact diagonalization of the position-space \sch equation~\cite{Fey2020}, a mixture of variational and perturbative approaches~\cite{Efimkin2021}, Fadeev equations~\cite{Mohseni2023}, and hyperspherical methods~\cite{Filikhin2018,Kezerashvili2024}.

\begin{figure}[t]
\begin{center}
\includegraphics[width=0.60\linewidth]{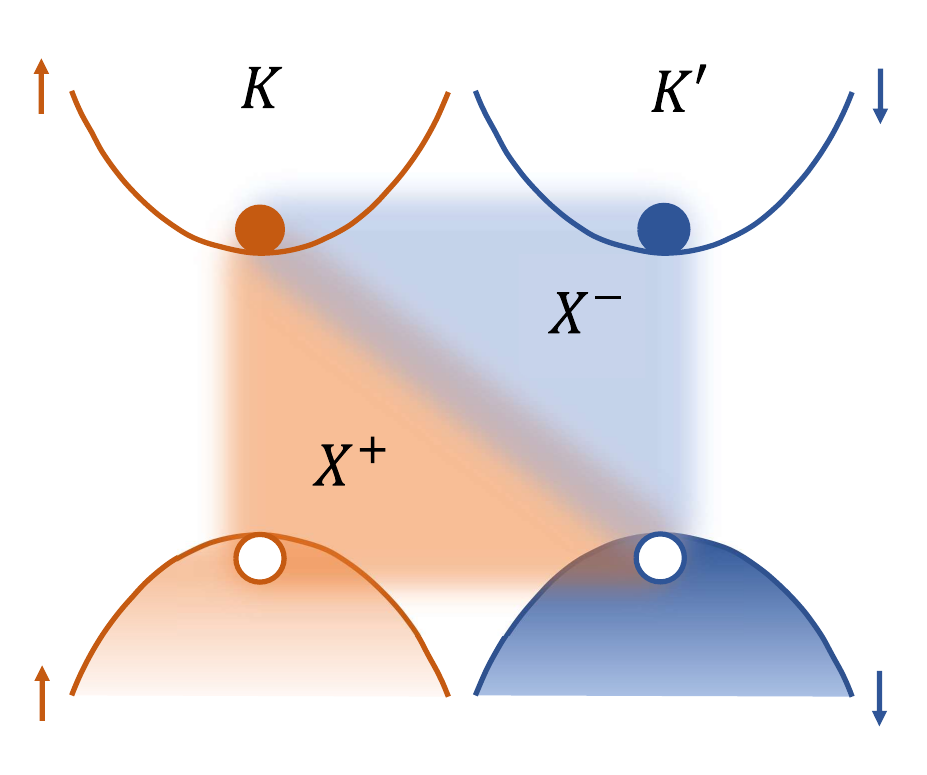}
\caption[system]{\label{fig:trion} 
Schematic illustration of distinguishable (or spin singlet) trions in monolayer TMDs, such as MoSe$_2$. Here the positively charged trion $X^+$ is formed by two holes in opposite valleys and one electron in either $K$ (orange shaded region) or $K'$ valley (not shown). Similarly, the negatively charged trion $X^-$ is formed by two electrons in opposite valleys and one hole in either $K$ (not shown) or $K'$ valley (blue shaded region).
}
\end{center}
\end{figure}

To be concrete, we consider trions consisting of distinguishable carriers, as illustrated in Fig.~\ref{fig:trion} for a MoSe$_2$ monolayer. However, our method can be straightforwardly applied also to the calculation of trions with, e.g., two identical electrons, which are expected to be present in systems with relatively large hole- to electron-mass ratio~\cite{Sergeev2001,Sergeev_Nanotechnology2001,Pricoupenko2010,Parish_PRA2013,Ngampruetikorn2013,Courtade2017,Tiene2022}. 

We begin by obtaining the trion Schr\"{o}dinger equation in momentum space. To this end, we introduce the (negatively charged) trion state $\ket{X^-} = \sum_{\vk\vp} \chi_{\vk,\vp} \ehd_{\up,\vk} \ehd_{\down,\vp}\hat{h}^{\dagger}_{\sigma,-\vk-\vp} \ket{0}$ where we explicitly consider distinguishable electrons. In terms of the trion wave function $\chi_{\vk,\vp}$, the \sch equation $\hat H\ket{X^-}=E\ket{X^-}$ takes the form
\begin{align}
&
\left(E-\epsilon^{e}_{\vk}-\epsilon^{e}_{\vp}-\epsilon^{h}_{\vk+\vp}\right) \chi_{\vk,\vp} = - \underset{\vk'}{\sum} V_{\vk-\vk'} \chi_{\vk',\vp} 
\nonumber \\
&
- \underset{\vp'}{\sum} V_{\vp-\vp'} \chi_{\vk,\vp'}
+ \underset{\vk'\vp'\vectorbold{Q}}{\sum} V_{\vectorbold{Q}} \delta_{\vk,\vk'+\vectorbold{Q}} \delta_{\vp,\vp'-\vectorbold{Q}} \chi_{\vk',\vp'} ,
\label{eq:TrionSEFull}
\end{align}
where $\vk$ and $\vp$ are the momenta of the electrons with respect to the hole. The first two terms on the right-hand side correspond to the attractive electron-hole interactions, and the last term to the repulsive electron-electron interaction, with $\vpd -\vp = \vk-\vkd \equiv \vQ$ being the momentum exchanged between the electrons. We note that the case of a positively charged trion can be obtained simply by exchanging electron and hole kinetic energies in Eq.~\eqref{eq:TrionSEFull}, i.e., by considering the transformation $m_e \leftrightarrow m_h$. The \sch equation has a straightforward diagrammatic interpretation, as shown in Fig.~\ref{TrionInteractionFig}. Note that this is distinct from the Feynman diagram treatment of the Coulomb three-body problem introduced in Ref.~\cite{CombescotPRX2017}. 

To carry out the subtraction scheme, it is convenient to rearrange the terms in Eq.~\eqref{eq:TrionSEFull} as follows:
\begin{align} \label{eq:TrionSE}
&
\left(E-\epsilon^{e}_{\vk}-\epsilon^{e}_{\vp}-\epsilon^{h}_{\vk+\vp}\right) \chi_{\vk,\vp} = \nonumber \\
&\hspace{2cm}- \underset{\vk'}{\sum} V_{\vk-\vk'} \left( \chi_{\vk',\vp} - \frac{1}{2}\chi_{\vk',\vk+\vp-\vk'} \right) \nonumber \\ 
&\hspace{2cm}
- \underset{\vp'}{\sum} V_{\vp-\vp'} \left( \chi_{\vk,\vp'} -\frac{1}{2} \chi_{\vk+\vp-\vp',\vp'}
\right) .
\end{align}
We then apply the Land{\'e} subtraction technique from Sec.~\ref{sec:Lande} to each term on the right-hand side to obtain
\begin{align} \label{eq:TrionSESubbed}
&
(E-\epsilon^{e}_{\vk}-\epsilon^{e}_{\vp}-\epsilon^{h}_{\vk+\vp}) \chi_{\vk,\vp} =
\nonumber \\
&
\hspace{2.1cm} - \underset{\vk'}{\sum} V_{\vk-\vk'} \bigg{(} \chi_{\vk',\vp} - \frac{1}{2}\chi_{\vk',\vk+\vp-\vk'} \bigg{)}
\nonumber \\
&
\hspace{1.825cm} + \bigg{(} \underset{\vk'}{\sum} g_{\vk,\vk'} - K_{k}   \bigg{)}
\frac{1}{2}\chi_{\vk,\vp}
\nonumber \\
&
\hspace{2.075cm} - \underset{\vp'}{\sum} V_{\vp-\vp'} \bigg{(} \chi_{\vk,\vp'} -\frac{1}{2} \chi_{\vk+\vp-\vp',\vp'} \bigg{)} 
\nonumber \\
&
\hspace{1.825cm} + \bigg{(}
\underset{\vp'}{\sum} g_{\vp,\vp'}  - K_{p}
\bigg{)}
\frac{1}{2} \chi_{\vk,\vp} .
\end{align}
This allows us to efficiently solve for the energy $E$ similarly to the exciton problem, Eq.~\eqref{eq:ExcitonSESubtracted}, and thus to obtain the trion binding energy $\varepsilon_T\equiv |E-E_{1s}|$. We note, however, that Eq.~\eqref{eq:TrionSESubbed} has momentum components $\vk+\vp-\vk'$ and $\vk+\vp-\vp'$, which are ``off-grid'' in the sense that $\chi$ on the right-hand side is evaluated at a point not necessarily on the momentum and angular grid specified on the left-hand side. This complicates the direct conversion of the equation into a solvable matrix form when evaluating on a Gaussian quadrature grid. We tackle this issue by the use of a Lanczos-type iterative method~\cite{Stadler1991,Demmel1997,Hadizadeh2007,Hadizadeh2012}, which we detail in Appendix~\ref{LanczosSect}, accompanied by an interpolation to evaluate the off-grid terms. Furthermore, all the data shown in the following is obtained by a linear extrapolation to infinite grid size---see Appendix~\ref{AppendixTMDCTable}.

\begin{figure}[th]
\begin{center}
\includegraphics[width=1.00\linewidth]{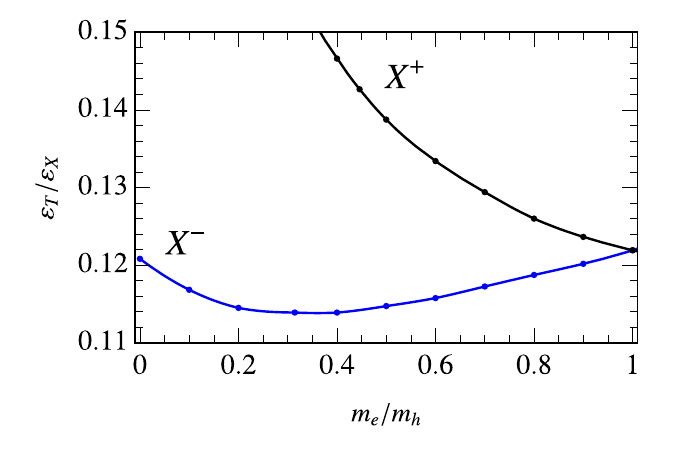}
\caption[system]{\label{fig:MassRatioTrion} 
Trion binding energies as a function of the electron-hole mass ratio in a conventional quantum well, where the electronic interactions are described by Eq.~\eqref{eq:Vcoulomb}. We show our results for the positive $X^+$ (black circles) and negative $X^-$ (blue circles) trions. Lines joining the data points are guides to the eye. Note that the $X^+$ line tends to a finite value in the limit $m_h/m_e \to \infty$~\cite{Stebe1989,Sergeev2001,Kidd_PRB2016,Szyniszewski2017}, but we terminate it at $m_h/m_e\simeq 2.5$ since our numerics converge too slowly beyond that point.}
\end{center}
\end{figure}

We first consider the case of the uniform Coulomb potential~\eqref{eq:Vcoulomb}, which describes the case of conventional quantum wells. We plot the trion binding energies for both the negatively and positively charged trions as a function of the electron-hole mass ratio in Fig.~\ref{fig:MassRatioTrion}. In particular we see that, in units of the exciton binding energy, the $X^-$ trion binding energy remains relatively constant, whereas the $X^+$ binding energy depends strongly on the mass ratio. A similar result has been obtained in Refs.~\cite{Usukura1999,Sergeev2001} from a variational wave function approach, although, being variational, this is expected to underestimate the binding energy as it provides an upper bound on the energy. Indeed, we find that our numerically exact binding energies are consistently larger than those in Ref.~\cite{Sergeev2001} (e.g., by about 10$\%$ for $m_e=m_h$). This illustrates that our exact method provides an improved description of the trion bound state compared with those assumed in the variational approaches of Refs.~\cite{Usukura1999,Sergeev2001}.

Of particular interest is the standard Coulomb case with $m_e=m_h$. Here, we find the trion binding energy $\varepsilon_T=0.122 \varepsilon_X$. Our calculation compares well with the Monte Carlo calculation of Ref.~\cite{Szyniszewski2017} which predicts the energy $\sim0.12\varepsilon_X$, and indicates a larger binding energy compared to the $\sim0.11\varepsilon_X$ predicted by calculations based on variational wave functions~\cite{Sergeev2001,Courtade2017}. 

\begin{figure}[th]
\begin{center}
\includegraphics[width=1.00\linewidth]{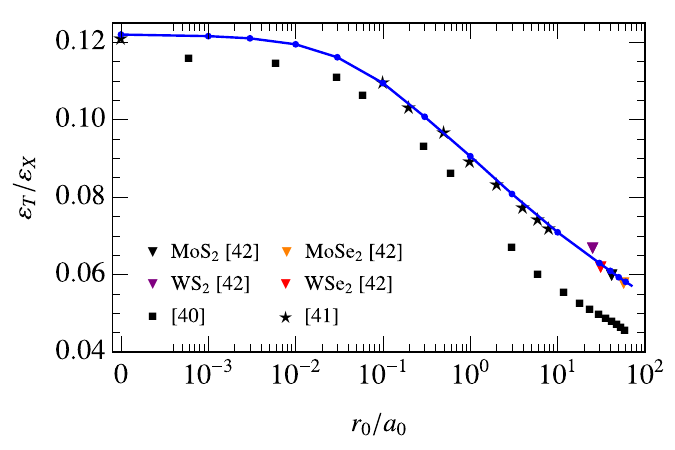}
\caption[system]{\label{fig:KeldyshTrion} 
Ratio of the trion binding energy $\varepsilon_T$ to the exciton binding energy $\varepsilon_X$ as a function of the screening length $r_0$ for $m_e=m_h$ (blue circles). We compare our results with those obtained from the variational calculations by Courtade et al.~\cite{Courtade2017} (black squares), from the Monte Carlo calculations by Szyniszewski et al.~\cite{Szyniszewski2017} (black stars), and from an exact diagonalization approach based on the discrete variable representation by Fey et al.~\cite{Fey2020} for different TMD material parameter values (triangles). The line joining our data points is a  guide to the eye.
}
\end{center}
\end{figure}

We now turn to the case of monolayer TMDs, i.e., where the electronic interactions are described by the Rytova-Keldysh potential of Eq.~\eqref{eq:VRK}. For simplicity, we start by considering the case of equal electron and hole masses, which is a reasonable approximation in TMDs~\cite{Rasmussen_PhysChemC2015}. We show the ratio between the trion and exciton binding energies as a function of the screening length $r_0$ in Fig.~\ref{fig:KeldyshTrion}. Interestingly, we see that this ratio only changes by roughly a factor two up to $r_0/a_0\simeq 50$ which is in the high end of realistic values of screening lengths for TMDs. Since this behavior does not reflect the sensitive dependence of the $1s$ exciton binding energy on the screening length, shown in Fig.~\ref{fig:Screenedfig:ExcitonFig}, this implies that the dependence of the trion binding energy must mirror that of the exciton. We also find that our calculated binding energies compare extremely well with the Monte Carlo calculations of Szyniszewski et al.~\cite{Szyniszewski2017}, and that they are up to about 25$\%$ larger than those obtained from variational wave functions by Courtade et al.~\cite{Courtade2017}. 

\begin{table*}
    \centering
    \renewcommand{\arraystretch}{1.5} 
    \renewcommand{\tabcolsep}{5.4pt}
\begin{tabular}{cccccccccc}
    \hline \hline
    TMD & $m_e (m_0)$ & $m_h (m_0)$ & $r_0 (\rm{\AA})$ & 
    $\varepsilon_X \text{(meV)}^*$ & 
    $\varepsilon_X \text{(meV)}$ & 
    $\varepsilon_T\text{(meV)}^*$  &
    $\varepsilon_T\text{(meV)}$  &
    ${\varepsilon_T/\varepsilon_X}^*$ &
    $\varepsilon_T/\varepsilon_X$  \\
    \hline 
    $\text{MoS\textsubscript{2}}$ & $0.47$ & $0.54$ & $44.6814$ & $526.0$ & $525.98$ & $31.7$ & $31.82$ & $0.0603$  & $0.0605$ \\
    $\text{MoSe\textsubscript{2}}$ & $0.55$ & $0.59$ & $53.1624$ & $476.7$ & $476.69$ & $27.7$ & $27.84$ & $0.0581$ & $0.0584$ \\
    $\text{WS\textsubscript{2}}$ & $0.32$ & $0.35$ & $40.1747$ & $508.6$ & $508.55$ & $34.2$ & $32.60$ & $0.0672$ & $0.0641$ \\
    $\text{WSe\textsubscript{2}}$ & $0.34$ & $0.36$ & $47.5701$ & $456.0$ & $456.02$ & $28.4$ & $28.55$ & $0.0623$ & $0.0626$  \\
    \hline \hline
\end{tabular}
    \caption{$1s$ exciton and ground state trion binding energies calculated for realistic material parameters for four different monolayer TMDs suspended in vacuum (i.e., for dielectric constant $\varepsilon=1)$. We compare our results with those of Ref.~\cite{Fey2020}, with the latter marked by an asterisk $(*)$. The material parameters used were obtained from density functional theory calculations in Ref.~\cite{Kylanpaa2015} which agree with experimental results in Ref.~\cite{Bjorkman2014}.} 
    \label{TMDTrionLargeTable}
\end{table*}

Our results in Fig.~\ref{fig:KeldyshTrion} also compare well with the results of an exact diagonalization method based on a discrete variable representation by Fey et al.~\cite{Fey2020} which considered specific TMD parameters---although the results from Ref.~\cite{Fey2020} include a slight mass imbalance, we find that this only marginally affects the comparison. To clearly demonstrate the excellent agreement, we have used the exact same TMD material parameters to obtain the comparison to Ref.~\cite{Fey2020} given in Table~\ref{TMDTrionLargeTable}.

Finally, we also compare with experimental measurements of the trion binding energy in MoSe$_2$, 
specifically for the case where all the theoretical input parameters, such as the dielectric screening, are well characterized~\cite{Ross2013,Berkelbach2013}. Here, our theory predicts a trion binding energy of 29 meV, which agrees well with the experimentally measured binding energy of 30 meV~\cite{Ross2013}.

The comparison of our results with previous calculations illustrate the power of combining the Land{\'e} subtraction scheme with a momentum-space representation of the problem of a few charged particles. Indeed, we have found that our trion binding energies are consistently larger than those obtained using variational methods, either based on wave functions~\cite{Sergeev2001,Courtade2017} or more advanced QMC calculations~\cite{Szyniszewski2017}. This
demonstrates that our method provides a promising approach for tackling a range of few- and many-body problems.

\section{Conclusion and outlook}
\label{sec:conc}

To conclude, we have calculated the binding energies of trions in 2D semiconductors using a numerical approach that applies equally to quantum wells such as III-V and II-VI structures and to TMD monolayers. Key to our results is a highly efficient way of treating the pole of the electronic interactions that occurs at small momentum exchange, thus allowing us to directly solve the \sch equation in momentum space. We have shown that our results are in excellent agreement with QMC calculations~\cite{Szyniszewski2017} and with recent exact diagonalization of the real-space \sch equation~\cite{Fey2020}, while yielding a larger binding energy than obtained in recent variational calculations~\cite{Courtade2017}.

Our numerical approach can be applied to a host of other few-body problems in 2D semiconductors, from larger complexes such as biexcitons~\cite{You-Heinz_NatPhys2015,Plechinger_PSS2015,Stevens_NatComm2018,Steinhoff-Li_NatPhys2018,Conway_2DMat2022} or charged biexcitons~\cite{Hao2017,Barbone-Atature_NatComm2018,Muir2022} to more exotic trions that involve either two identical electrons (or holes), electrons belonging to different bands and thus with different parabolic dispersions, trions in anisotropic materials such as phosphorene~\cite{Yang2015phosphorene,Xu2016phosphorene}, or even charges from different layers in heterostructures. It could also be extended to the study of scattering problems, for instance to a fully microscopic approach to electron--Rydberg-exciton scattering or to electron--exciton-polariton scattering~\cite{Li2021PRL,Kumar2023}. Another avenue of interest is to include electron-hole exchange processes, which is a necessary ingredient in describing the difference between spin triplet and singlet trions in tungsten-based TMDs~\cite{Plechinger2016}.

Finally, our approach can also potentially be applied to many-body problems that involve excitons and electrons. A prominent such problem is the exciton Fermi polaron~\cite{SidlerNatPhys16}, where an exciton is immersed in an electron Fermi sea. In the regime of low to intermediate doping, the system is quite well understood~\cite{SidlerNatPhys16,Efimkin2017,Huang2023} in terms of a tightly bound exciton dressed by electrons which results in a ground state attractive polaron that is continuously connected to the trion. However, at higher doping there are several puzzling features in the optical response~\cite{Jones2013,Li2021Exciton}, and their detailed theoretical modelling may require one to explicitly treat the exciton's constituent particles. Our numerical approach is ideally suited to treat such a problem. Thus, the efficient description of electronic interactions presented here may have benefits throughout the full range of few- to many-body problems.

\begin{acknowledgments}
We acknowledge Emma Laird for discussions on how to implement the Land{\'e} subtraction technique, and Dmitry Efimkin for discussions on excitons and trions. We thank M.~Szyniszewski for sharing data from Ref.~\cite{Szyniszewski2017} and M.~M.~Glazov and M.~Semina for sharing data from Ref.~\cite{Courtade2017}.  We acknowledge support from the Australian Research Council Centre of Excellence in Future Low Energy Electronics Technologies (CE170100039). SSK is supported by the Monash University Postgraduate Publication Award. JL and MMP are also supported by the Australian Research Council Future Fellowships FT160100244 and FT200100619, respectively, and JL is supported by Australian Research Council Discovery Project DP240100569. This research was supported in part by the Monash eResearch Centre and eSolutions-Research Support Services through the use of the MonARCH HPC Cluster. FMM acknowledges financial support from the Spanish Ministry of Science, Innovation and Universities through the ``Maria de Maetzu'' Programme for Units of Excellence in R\&D (CEX2023-001316-M), from the Ministry of Science, Innovation and Universities MCIN/AEI/10.13039/501100011033, FEDER UE,  projects No.~PID2020-113415RB-C22 (2DEnLight), and No.~PID2023-150420NB-C31 (Q), and from the Proyecto Sinérgico CAM 2020 Y2020/TCS-6545 (NanoQuCo-CM).
\end{acknowledgments}

\appendix

\section{Lanczos-type algorithm for the eigenenergy} \label{LanczosSect}

\begin{figure}[ht]
\begin{center}
\includegraphics[width=.95\linewidth]{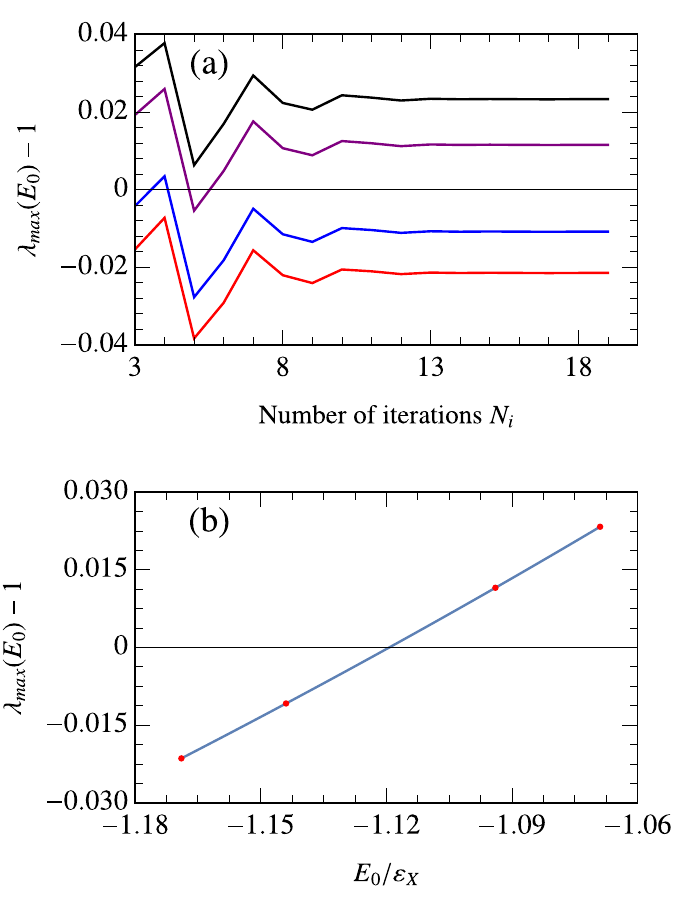}
\caption[system]{\label{AppendixLanczosConvergence}
(a) The value of $\lambda_{max} (E_0) - 1$ in the Lanczos algorithm applied to the equal mass Coulomb trion \sch equation~\eqref{eq:TrionSE} as a function of the number of iterations $N_i$. The red, blue, purple, and black lines correspond to the initial energy estimates of $E_0/\varepsilon_X=$ -1.16893, -1.14393, -1.09393, and -1.06893, respectively. (b) The converged $\lambda_{max} (E_0) - 1$ values and the corresponding linear interpolation are shown, which is used to calculate the eigenenergy $E/\varepsilon_X=-1.11925$ at $\lambda_{max} \rightarrow 1$. The momentum and angular grid sizes are $N_k=40$ and $N_\theta=20$, and the grid scaling factor $m=1.1$.
}
\end{center}
\end{figure}

Here, we go through the Lanczos-type algorithm used in Refs.~\cite{Stadler1991,Demmel1997,Hadizadeh2007,Hadizadeh2012} to find the eigenenergy by the method of iteration. In brief, the iterative method uses repeated operations of Eq.~\eqref{eq:TrionSE} starting from an arbitrary initial guess $\chi_0$ for the wave function, from which the physical eigenenergy and wave function can be found. This procedure can be performed without transforming the whole equation into a matrix form. The $\chi_1$ produced from operating on $\chi_0$ is interpolated, which is required for the next iteration to produce $\chi_2$. The iteration is repeated and the resulting set of $\{\chi_i\}$ is used to calculate the eigenenergy.

Specifically, an eigenvalue integral equation such as Eq.~\eqref{eq:ExcitonSE} or Eq.~\eqref{eq:TrionSEFull} can be written in operator form as
\begin{equation}
\lambda(E) \ket{\psi} = K (E) \ket{ \psi }  , 
\end{equation}
where $K$ represents the operation performed on the eigenfunction/wave function $\ket{\psi}$, and where we arrange the equation such that the numerically exact energy $E$ 
corresponds to $\lambda (E)=1$. For instance, in Eq.~\eqref{eq:TrionSESubbed} $K$ is the operator acting on $\chi_{\vk,\vp}$ after dividing both sides of the equation by $E-\epsilon^{e}_{\vk}-\epsilon^{e}_{\vp}-\epsilon^{h}_{\vk+\vp}$.

We begin by using an initial guess $\ket{\psi_0}$ for $\ket{\psi}$, which roughly resembles the actual wave function, and similarly, an estimated energy $E_0$. We then repeatedly apply the operation $K$ 
\begin{equation}
\ket{\psi_{i+1}} = K(E_0) \ket{\psi_i},
\end{equation}
$N$ times, from which we obtain the set of vectors $\{\ket{\psi_i}\}$. 

An orthogonal basis of vectors $\{ \ket{\bar{\psi_i}} \}$ is constructed from $\{ \ket{\psi_i} \}$ using any algorithm of choice, such as the Gram–Schmidt algorithm. The two sets of vectors can be represented in terms of each other as
\begin{subequations}
\begin{align}
&
\ket{\psi_i}=\underset{j=0}{\overset{i}{\sum}} a_{ij}\ket{\bar{\psi}_j} ,
\\
&
\ket{\bar{\psi}_i}=\underset{j=0}{\overset{i}{\sum}} b_{ij} \ket{\psi_j} ,
\end{align}    
\end{subequations}
where the coefficients $a_{ij}$ and $b_{ij}$ can be found by using orthogonality and dot products. 

The eigenvector $\ket{\psi}$ can then be expanded in the orthonormal basis
\begin{equation}
\ket{\psi} = \underset{i=0}{\overset{N}{\sum}} c_i \ket{\bar{\psi}_i} .    
\end{equation}
We now substitute this expansion into the original eigenequation to obtain,
\begin{align}
\underset{i=0}{\overset{N}{\sum}} c_i K(E_0) \ket{\bar{\psi}_i} 
& 
= \underset{i=0}{\overset{N}{\sum}} c_i \underset{j=0}{\overset{i}{\sum}} b_{ij} K(E_0) \ket{\psi_j} \nn
\\
&
= \underset{i=0}{\overset{N}{\sum}} c_i \underset{j=0}{\overset{i}{\sum}} b_{ij} \ket{\psi_{j+1}} \nn
\\
&
= \underset{i=0}{\overset{N}{\sum}} c_i \underset{j=0}{\overset{i}{\sum}} b_{ij}  \underset{k=0}{\overset{j+1}{\sum}} a_{j+1,k} \ket{\bar{\psi}_{k}} \nn
\\
&
=\lambda(E_0) \underset{i=0}{\overset{N}{\sum}} c_i \ket{\bar{\psi}_i}
\end{align}
Multiplying by $\bra{\bar{\psi}_n}$ from the left, we obtain
\begin{align}
\underset{i=0}{\overset{N}{\sum}} c_i M_{in} &
= \lambda(E_0) c_n, 
\end{align}
where
\begin{align}
M_{in} 
&
= \underset{j=0}{\overset{i}{\sum}} b_{ij} a_{j+1,n} ,
\end{align}
for $0<n\leq N$. In other words, the eigenvalues of the matrix $M$ correspond to precisely $\lambda(E_0)$. Now, the physical eigenenergy $E$ can be found for which $\lambda_{max}(E)\rightarrow1$, where $\lambda_{max} (E)$ is the largest of the eigenvalues $\lambda(E)$, which dominates the iteration. 

\begin{figure}[th]
\begin{center}
\includegraphics[width=.95\linewidth]{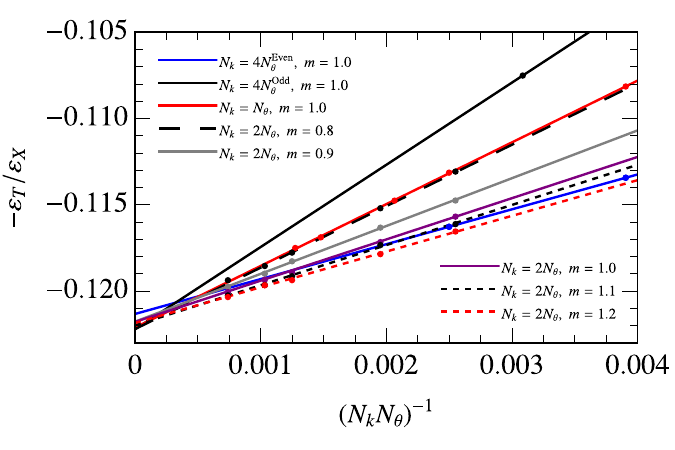}
\caption[system]{\label{TrionConvergence} 
Trion binding energy for $m_e=m_h$ and $r_0=0$ as a function of different grid sizes, where $N_k$ and $N_\theta$ are the number of momentum and angular grid points, respectively. The linear fit for different grid schemes and scaling factors $m$ are shown, along with calculated values (circles). Some data points used for the linear fits are outside the range shown.}
\end{center}
\end{figure}

\begin{table*}[t]
    \centering
    \renewcommand{\arraystretch}{1} 
    \setlength{\tabcolsep}{10pt} 
    {%
    \begin{tabular}{|c|c|c|c|c|c|c|c|c|c|}
    \hline
    \multirow{2}{*}{TMD} & \multirow{2}{*}{$m_e (m_0)$} & \multirow{2}{*}{$m_h (m_0)$} & \multirow{2}{*}{$r_0 (\rm{\AA})$} & \multicolumn{6}{c|}{$\varepsilon_T/\varepsilon_X$} \\
    \cline{5-10}
    & & & & $N_k=28$ & $32$ & $40$ & $44$ & $52$ & $\infty$ \\
    \hline
    $\text{MoS\textsubscript{2}}$ & $0.47$ & $0.54$ & $44.6814$ & $0.05582$ & $0.05690$ & $0.05822$ & $0.05858$ & $0.05918$ & $0.06052$ \\
    \hline
    $\text{MoSe\textsubscript{2}}$ & $0.55$ & $0.59$ & $53.1624$ & $0.05381$ & $0.05486$ & $0.05619$ & $0.05661$ & $0.05706$ & $0.05845$  \\
    \hline
    $\text{WS\textsubscript{2}}$ & $0.32$ & $0.35$ & $40.1747$ & $0.05931$ & $0.06045$ & $0.06184$ & $0.06216$ & $0.06274$ & $0.06415$ \\
    \hline
    $\text{WSe\textsubscript{2}}$ & $0.34$ & $0.36$ & $47.5701$ & $0.05781$ & $0.05898$ & $0.06024$ & $0.06069$ & $0.06125$ & $0.06265$ \\
    \hline
    \end{tabular}
    }
    \caption{
    The ratio of the screened trion binding energy $\varepsilon_T$ to the screened exciton energy $\varepsilon_X$ calculated using the Land{\'e} subtraction technique is shown for TMD material parameters. The calculated values of $\varepsilon_T/\varepsilon_X$ are tabulated for $N_k=2N_\theta$ grid with scaling $m=1$, for $N_k=$ 28, 32, 40, 44, 52, along with the extrapolated infinite grid values.
    } 
    \label{AppendixTMDTrionTable}
\end{table*}

An advantage of the Lanczos algorithm is its iterative nature, allowing sequential execution with summations or vector multiplications without requiring an explicit kernel matrix.
To apply this method to the trion, we use the free propagator of the system $G_0 (\vk,\vp,E) =  \frac{1}{E-\epsilon^{e}_{\vk}-\epsilon^{e}_{\vp}-\epsilon^{h}_{\vk+\vp}}$ as the initial guess $\chi_0$ in solving Eq.~\eqref{eq:TrionSESubbed}, and find the convergence of $\lambda_{max} (E_0)$ to be quite independent of the exact functional form. 
We now substitute the momentum and angular quadrature points into Eq.~\eqref{eq:TrionSESubbed} and calculate the first element of $\chi_1$ (corresponding to the first momentum and angular quadrature point) by performing the integrations numerically over the initial guess $\chi_0$, and this process can be repeated to calculate $\chi_{1: \vk,\vp}$ over all quadrature points.
We also use a 2D linear interpolation/extrapolation for $\chi_i$ for $i>0$ after transforming the momentum to a logarithmic scale, in order to evaluate the off-grid points used to calculate $\chi_{i+1}$. 

The Gauss-Legendre quadrature points $z\in (-1,1)$ are transformed to the full momentum range $k\in (0,\infty)$ using the transformation $(\frac{2}{1+z}-1)^{1/m}$, and within the numerical solution the momentum was measured in units of $\sqrt{2m_r\varepsilon_X}$ to improve convergence. The angular grid points were chosen to be equally spaced from $0$ to $2\pi$, corresponding to a Riemann sum. We perform the calculation for a different number of momentum and angular grid points, $N_k$ and $N_\theta$, respectively.

In Fig.~\ref{AppendixLanczosConvergence}, we illustrate this method for the case of the trion energy with $r_0=0$ and $m_e=m_h$, using $N_k=40$ and $N_\theta=20$ and a grid scaling of $m=1.1$. Figure~\ref{AppendixLanczosConvergence}(a) shows $1-\lambda_{max} (E_0)$ as a function of the number of iterations $N_i\leq N$. The initial energy estimates $E_0$ and their converged $\lambda_{max} (E_0) - 1$ values are shown in Fig.~\ref{AppendixLanczosConvergence}(b), along with the linear interpolation used to find the eigenenergy $E$ corresponding to $\lambda_{max} \rightarrow 1$.

Computationally, we allocated 1 CPU core per energy input, with no parallelization used internally. One equal mass Coulomb calculation with $\{N_k, N_{\theta}\} = \{32, 16\}$ took approximately 8 hours with 110GB memory allocated and $\{N_k, N_{\theta}\} = \{52, 26\}$ took 160 hours with 190GB memory allocated. Calculations with screening generally took less time to compute than those without.

\section{Extrapolation to infinite grid size} \label{AppendixTMDCTable}

The results shown in the main text are obtained by extrapolating to infinite grid size. To illustrate the procedure used, in Fig.~\ref{TrionConvergence} we plot the trion binding energy for the specific case of $m_e=m_h$ and $r_0=0$ as a function of the inverse grid size ($1/ N_k N_\theta$) for different ratios of $N_k$, $N_\theta$, and grid scaling $m$ (see Appendix~\ref{LanczosSect}). Independently of the precise details of the ratio of $N_k$ to $N_\theta$ and the precise value of $m$, the energy is observed to vary linearly with the inverse grid size. This allows for linear fitting, and extrapolation of the energy to a grid of infinite size. For instance, in this case the trion binding energy can be seen to converge to $\sim 0.122\varepsilon_X$ at infinite grid size, independent of the grid's specific details.

Specifically, the results shown in the main text are obtained by extrapolating to infinite grid size for $N_k=2N_\theta$ with grid scaling $m=1$. This includes the data in Figs.~\ref{fig:MassRatioTrion} and \ref{fig:KeldyshTrion} and also that in Table~\ref{TMDTrionLargeTable}. 

To further illustrate this procedure, in Table~\ref{AppendixTMDTrionTable} we show the data points used to extrapolate the trion energies for the specific TMD values in Table \ref{TMDTrionLargeTable} to an infinite grid size. This again shows a very clear linear dependence.

\bibliography{trionPRB_refs} 

\begin{thebibliography}{88}%
\makeatletter
\providecommand \@ifxundefined [1]{%
 \@ifx{#1\undefined}
}%
\providecommand \@ifnum [1]{%
 \ifnum #1\expandafter \@firstoftwo
 \else \expandafter \@secondoftwo
 \fi
}%
\providecommand \@ifx [1]{%
 \ifx #1\expandafter \@firstoftwo
 \else \expandafter \@secondoftwo
 \fi
}%
\providecommand \natexlab [1]{#1}%
\providecommand \enquote  [1]{``#1''}%
\providecommand \bibnamefont  [1]{#1}%
\providecommand \bibfnamefont [1]{#1}%
\providecommand \citenamefont [1]{#1}%
\providecommand \href@noop [0]{\@secondoftwo}%
\providecommand \href [0]{\begingroup \@sanitize@url \@href}%
\providecommand \@href[1]{\@@startlink{#1}\@@href}%
\providecommand \@@href[1]{\endgroup#1\@@endlink}%
\providecommand \@sanitize@url [0]{\catcode `\\12\catcode `\$12\catcode
  `\&12\catcode `\#12\catcode `\^12\catcode `\_12\catcode `\%12\relax}%
\providecommand \@@startlink[1]{}%
\providecommand \@@endlink[0]{}%
\providecommand \url  [0]{\begingroup\@sanitize@url \@url }%
\providecommand \@url [1]{\endgroup\@href {#1}{\urlprefix }}%
\providecommand \urlprefix  [0]{URL }%
\providecommand \Eprint [0]{\href }%
\providecommand \doibase [0]{http://dx.doi.org/}%
\providecommand \selectlanguage [0]{\@gobble}%
\providecommand \bibinfo  [0]{\@secondoftwo}%
\providecommand \bibfield  [0]{\@secondoftwo}%
\providecommand \translation [1]{[#1]}%
\providecommand \BibitemOpen [0]{}%
\providecommand \bibitemStop [0]{}%
\providecommand \bibitemNoStop [0]{.\EOS\space}%
\providecommand \EOS [0]{\spacefactor3000\relax}%
\providecommand \BibitemShut  [1]{\csname bibitem#1\endcsname}%
\let\auto@bib@innerbib\@empty
\bibitem [{\citenamefont {Britnell}\ \emph {et~al.}(2013)\citenamefont
  {Britnell}, \citenamefont {Ribeiro}, \citenamefont {Eckmann}, \citenamefont
  {Jalil}, \citenamefont {Belle}, \citenamefont {Mishchenko}, \citenamefont
  {Kim}, \citenamefont {Gorbachev}, \citenamefont {Georgiou}, \citenamefont
  {Morozov}, \citenamefont {Grigorenko}, \citenamefont {Geim}, \citenamefont
  {Casiraghi}, \citenamefont {Neto},\ and\ \citenamefont
  {Novoselov}}]{Britnell2013}%
  \BibitemOpen
  \bibfield  {author} {\bibinfo {author} {\bibfnamefont {L.}~\bibnamefont
  {Britnell}}, \bibinfo {author} {\bibfnamefont {R.~M.}\ \bibnamefont
  {Ribeiro}}, \bibinfo {author} {\bibfnamefont {A.}~\bibnamefont {Eckmann}},
  \bibinfo {author} {\bibfnamefont {R.}~\bibnamefont {Jalil}}, \bibinfo
  {author} {\bibfnamefont {B.~D.}\ \bibnamefont {Belle}}, \bibinfo {author}
  {\bibfnamefont {A.}~\bibnamefont {Mishchenko}}, \bibinfo {author}
  {\bibfnamefont {Y.-J.}\ \bibnamefont {Kim}}, \bibinfo {author} {\bibfnamefont
  {R.~V.}\ \bibnamefont {Gorbachev}}, \bibinfo {author} {\bibfnamefont
  {T.}~\bibnamefont {Georgiou}}, \bibinfo {author} {\bibfnamefont {S.~V.}\
  \bibnamefont {Morozov}}, \bibinfo {author} {\bibfnamefont {A.~N.}\
  \bibnamefont {Grigorenko}}, \bibinfo {author} {\bibfnamefont {A.~K.}\
  \bibnamefont {Geim}}, \bibinfo {author} {\bibfnamefont {C.}~\bibnamefont
  {Casiraghi}}, \bibinfo {author} {\bibfnamefont {A.~H.~C.}\ \bibnamefont
  {Neto}}, \ and\ \bibinfo {author} {\bibfnamefont {K.~S.}\ \bibnamefont
  {Novoselov}},\ }\bibfield  {title} {\bibinfo {title} {\emph {Strong
  {Light}-{Matter} {Interactions} in {Heterostructures} of {Atomically} {Thin}
  {Films}}},\ }\href {\doibase 10.1126/science.1235547} {\bibfield  {journal}
  {\bibinfo  {journal} {Science}\ }\textbf {\bibinfo {volume} {340}},\ \bibinfo
  {pages} {1311} (\bibinfo {year} {2013})}\BibitemShut {NoStop}%
\bibitem [{\citenamefont {Schaibley}\ \emph {et~al.}(2016)\citenamefont
  {Schaibley}, \citenamefont {Yu}, \citenamefont {Clark}, \citenamefont
  {Rivera}, \citenamefont {Ross}, \citenamefont {Seyler}, \citenamefont {Yao},\
  and\ \citenamefont {Xu}}]{Schaibley2016}%
  \BibitemOpen
  \bibfield  {author} {\bibinfo {author} {\bibfnamefont {J.~R.}\ \bibnamefont
  {Schaibley}}, \bibinfo {author} {\bibfnamefont {H.}~\bibnamefont {Yu}},
  \bibinfo {author} {\bibfnamefont {G.}~\bibnamefont {Clark}}, \bibinfo
  {author} {\bibfnamefont {P.}~\bibnamefont {Rivera}}, \bibinfo {author}
  {\bibfnamefont {J.~S.}\ \bibnamefont {Ross}}, \bibinfo {author}
  {\bibfnamefont {K.~L.}\ \bibnamefont {Seyler}}, \bibinfo {author}
  {\bibfnamefont {W.}~\bibnamefont {Yao}}, \ and\ \bibinfo {author}
  {\bibfnamefont {X.}~\bibnamefont {Xu}},\ }\bibfield  {title} {\bibinfo
  {title} {\emph {Valleytronics in {2D} materials}},\ }\href {\doibase
  10.1038/natrevmats.2016.55} {\bibfield  {journal} {\bibinfo  {journal}
  {Nature Reviews Materials}\ }\textbf {\bibinfo {volume} {1}},\ \bibinfo
  {pages} {16055} (\bibinfo {year} {2016})}\BibitemShut {NoStop}%
\bibitem [{\citenamefont {Velický}\ and\ \citenamefont
  {Toth}(2017)}]{Velicky2017}%
  \BibitemOpen
  \bibfield  {author} {\bibinfo {author} {\bibfnamefont {M.}~\bibnamefont
  {Velický}}\ and\ \bibinfo {author} {\bibfnamefont {P.~S.}\ \bibnamefont
  {Toth}},\ }\bibfield  {title} {\bibinfo {title} {\emph {From two-dimensional
  materials to their heterostructures: An electrochemist's perspective}},\
  }\href {\doibase https://doi.org/10.1016/j.apmt.2017.05.003} {\bibfield
  {journal} {\bibinfo  {journal} {Applied Materials Today}\ }\textbf {\bibinfo
  {volume} {8}},\ \bibinfo {pages} {68} (\bibinfo {year} {2017})},\ \bibinfo
  {note} {2D Materials in Electrochemistry}\BibitemShut {NoStop}%
\bibitem [{\citenamefont {He}\ \emph {et~al.}(2014)\citenamefont {He},
  \citenamefont {Kumar}, \citenamefont {Zhao}, \citenamefont {Wang},
  \citenamefont {Mak}, \citenamefont {Zhao},\ and\ \citenamefont
  {Shan}}]{He2014}%
  \BibitemOpen
  \bibfield  {author} {\bibinfo {author} {\bibfnamefont {K.}~\bibnamefont
  {He}}, \bibinfo {author} {\bibfnamefont {N.}~\bibnamefont {Kumar}}, \bibinfo
  {author} {\bibfnamefont {L.}~\bibnamefont {Zhao}}, \bibinfo {author}
  {\bibfnamefont {Z.}~\bibnamefont {Wang}}, \bibinfo {author} {\bibfnamefont
  {K.~F.}\ \bibnamefont {Mak}}, \bibinfo {author} {\bibfnamefont
  {H.}~\bibnamefont {Zhao}}, \ and\ \bibinfo {author} {\bibfnamefont
  {J.}~\bibnamefont {Shan}},\ }\bibfield  {title} {\bibinfo {title} {\emph
  {Tightly Bound Excitons in Monolayer ${\mathrm{WSe}}_{2}$}},\ }\href
  {\doibase 10.1103/PhysRevLett.113.026803} {\bibfield  {journal} {\bibinfo
  {journal} {Phys. Rev. Lett.}\ }\textbf {\bibinfo {volume} {113}},\ \bibinfo
  {pages} {026803} (\bibinfo {year} {2014})}\BibitemShut {NoStop}%
\bibitem [{\citenamefont {Chernikov}\ \emph {et~al.}(2014)\citenamefont
  {Chernikov}, \citenamefont {Berkelbach}, \citenamefont {Hill}, \citenamefont
  {Rigosi}, \citenamefont {Li}, \citenamefont {Aslan}, \citenamefont
  {Reichman}, \citenamefont {Hybertsen},\ and\ \citenamefont
  {Heinz}}]{Chernikov2014}%
  \BibitemOpen
  \bibfield  {author} {\bibinfo {author} {\bibfnamefont {A.}~\bibnamefont
  {Chernikov}}, \bibinfo {author} {\bibfnamefont {T.~C.}\ \bibnamefont
  {Berkelbach}}, \bibinfo {author} {\bibfnamefont {H.~M.}\ \bibnamefont
  {Hill}}, \bibinfo {author} {\bibfnamefont {A.}~\bibnamefont {Rigosi}},
  \bibinfo {author} {\bibfnamefont {Y.}~\bibnamefont {Li}}, \bibinfo {author}
  {\bibfnamefont {B.}~\bibnamefont {Aslan}}, \bibinfo {author} {\bibfnamefont
  {D.~R.}\ \bibnamefont {Reichman}}, \bibinfo {author} {\bibfnamefont {M.~S.}\
  \bibnamefont {Hybertsen}}, \ and\ \bibinfo {author} {\bibfnamefont {T.~F.}\
  \bibnamefont {Heinz}},\ }\bibfield  {title} {\bibinfo {title} {\emph {Exciton
  Binding Energy and Nonhydrogenic Rydberg Series in Monolayer
  ${\mathrm{WS}}_{2}$}},\ }\href {\doibase 10.1103/PhysRevLett.113.076802}
  {\bibfield  {journal} {\bibinfo  {journal} {Phys. Rev. Lett.}\ }\textbf
  {\bibinfo {volume} {113}},\ \bibinfo {pages} {076802} (\bibinfo {year}
  {2014})}\BibitemShut {NoStop}%
\bibitem [{\citenamefont {Mak}\ \emph {et~al.}(2013)\citenamefont {Mak},
  \citenamefont {He}, \citenamefont {Lee}, \citenamefont {Lee}, \citenamefont
  {Hone}, \citenamefont {Heinz},\ and\ \citenamefont {Shan}}]{Mak2013}%
  \BibitemOpen
  \bibfield  {author} {\bibinfo {author} {\bibfnamefont {K.~F.}\ \bibnamefont
  {Mak}}, \bibinfo {author} {\bibfnamefont {K.}~\bibnamefont {He}}, \bibinfo
  {author} {\bibfnamefont {C.}~\bibnamefont {Lee}}, \bibinfo {author}
  {\bibfnamefont {G.~H.}\ \bibnamefont {Lee}}, \bibinfo {author} {\bibfnamefont
  {J.}~\bibnamefont {Hone}}, \bibinfo {author} {\bibfnamefont {T.~F.}\
  \bibnamefont {Heinz}}, \ and\ \bibinfo {author} {\bibfnamefont
  {J.}~\bibnamefont {Shan}},\ }\bibfield  {title} {\bibinfo {title} {\emph
  {Tightly bound trions in monolayer {MoS2}}},\ }\href {\doibase
  10.1038/nmat3505} {\bibfield  {journal} {\bibinfo  {journal} {Nature
  Materials}\ }\textbf {\bibinfo {volume} {12}},\ \bibinfo {pages} {207}
  (\bibinfo {year} {2013})}\BibitemShut {NoStop}%
\bibitem [{\citenamefont {Ross}\ \emph {et~al.}(2013)\citenamefont {Ross},
  \citenamefont {Wu}, \citenamefont {Yu}, \citenamefont {Ghimire},
  \citenamefont {Jones}, \citenamefont {Aivazian}, \citenamefont {Yan},
  \citenamefont {Mandrus}, \citenamefont {Xiao}, \citenamefont {Yao},\ and\
  \citenamefont {Xu}}]{Ross2013}%
  \BibitemOpen
  \bibfield  {author} {\bibinfo {author} {\bibfnamefont {J.~S.}\ \bibnamefont
  {Ross}}, \bibinfo {author} {\bibfnamefont {S.}~\bibnamefont {Wu}}, \bibinfo
  {author} {\bibfnamefont {H.}~\bibnamefont {Yu}}, \bibinfo {author}
  {\bibfnamefont {N.~J.}\ \bibnamefont {Ghimire}}, \bibinfo {author}
  {\bibfnamefont {A.~M.}\ \bibnamefont {Jones}}, \bibinfo {author}
  {\bibfnamefont {G.}~\bibnamefont {Aivazian}}, \bibinfo {author}
  {\bibfnamefont {J.}~\bibnamefont {Yan}}, \bibinfo {author} {\bibfnamefont
  {D.~G.}\ \bibnamefont {Mandrus}}, \bibinfo {author} {\bibfnamefont
  {D.}~\bibnamefont {Xiao}}, \bibinfo {author} {\bibfnamefont {W.}~\bibnamefont
  {Yao}}, \ and\ \bibinfo {author} {\bibfnamefont {X.}~\bibnamefont {Xu}},\
  }\bibfield  {title} {\bibinfo {title} {\emph {Electrical control of neutral
  and charged excitons in a monolayer semiconductor}},\ }\href {\doibase
  10.1038/ncomms2498} {\bibfield  {journal} {\bibinfo  {journal} {Nature
  Communications}\ }\textbf {\bibinfo {volume} {4}},\ \bibinfo {pages} {1474}
  (\bibinfo {year} {2013})}\BibitemShut {NoStop}%
\bibitem [{\citenamefont {Wang}\ \emph {et~al.}(2014)\citenamefont {Wang},
  \citenamefont {Bouet}, \citenamefont {Lagarde}, \citenamefont {Vidal},
  \citenamefont {Balocchi}, \citenamefont {Amand}, \citenamefont {Marie},\ and\
  \citenamefont {Urbaszek}}]{Wang-Urbaszek_PRB2014}%
  \BibitemOpen
  \bibfield  {author} {\bibinfo {author} {\bibfnamefont {G.}~\bibnamefont
  {Wang}}, \bibinfo {author} {\bibfnamefont {L.}~\bibnamefont {Bouet}},
  \bibinfo {author} {\bibfnamefont {D.}~\bibnamefont {Lagarde}}, \bibinfo
  {author} {\bibfnamefont {M.}~\bibnamefont {Vidal}}, \bibinfo {author}
  {\bibfnamefont {A.}~\bibnamefont {Balocchi}}, \bibinfo {author}
  {\bibfnamefont {T.}~\bibnamefont {Amand}}, \bibinfo {author} {\bibfnamefont
  {X.}~\bibnamefont {Marie}}, \ and\ \bibinfo {author} {\bibfnamefont
  {B.}~\bibnamefont {Urbaszek}},\ }\bibfield  {title} {\bibinfo {title} {\emph
  {Valley dynamics probed through charged and neutral exciton emission in
  monolayer ${\mathrm{WSe}}_{2}$}},\ }\href {\doibase
  10.1103/PhysRevB.90.075413} {\bibfield  {journal} {\bibinfo  {journal} {Phys.
  Rev. B}\ }\textbf {\bibinfo {volume} {90}},\ \bibinfo {pages} {075413}
  (\bibinfo {year} {2014})}\BibitemShut {NoStop}%
\bibitem [{\citenamefont {Zhu}\ \emph {et~al.}(2015)\citenamefont {Zhu},
  \citenamefont {Chen},\ and\ \citenamefont {Cui}}]{BZhu2015}%
  \BibitemOpen
  \bibfield  {author} {\bibinfo {author} {\bibfnamefont {B.}~\bibnamefont
  {Zhu}}, \bibinfo {author} {\bibfnamefont {X.}~\bibnamefont {Chen}}, \ and\
  \bibinfo {author} {\bibfnamefont {X.}~\bibnamefont {Cui}},\ }\bibfield
  {title} {\bibinfo {title} {\emph {Exciton {Binding} {Energy} of {Monolayer}
  {WS2}}},\ }\href {\doibase 10.1038/srep09218} {\bibfield  {journal} {\bibinfo
   {journal} {Scientific Reports}\ }\textbf {\bibinfo {volume} {5}},\ \bibinfo
  {pages} {9218} (\bibinfo {year} {2015})}\BibitemShut {NoStop}%
\bibitem [{\citenamefont {Yang}\ \emph
  {et~al.}(2015{\natexlab{a}})\citenamefont {Yang}, \citenamefont {Lü},
  \citenamefont {Myint}, \citenamefont {Pei}, \citenamefont {Macdonald},
  \citenamefont {Zheng},\ and\ \citenamefont {Lu}}]{Yang2015}%
  \BibitemOpen
  \bibfield  {author} {\bibinfo {author} {\bibfnamefont {J.}~\bibnamefont
  {Yang}}, \bibinfo {author} {\bibfnamefont {T.}~\bibnamefont {Lü}}, \bibinfo
  {author} {\bibfnamefont {Y.~W.}\ \bibnamefont {Myint}}, \bibinfo {author}
  {\bibfnamefont {J.}~\bibnamefont {Pei}}, \bibinfo {author} {\bibfnamefont
  {D.}~\bibnamefont {Macdonald}}, \bibinfo {author} {\bibfnamefont {J.-C.}\
  \bibnamefont {Zheng}}, \ and\ \bibinfo {author} {\bibfnamefont
  {Y.}~\bibnamefont {Lu}},\ }\bibfield  {title} {\bibinfo {title} {\emph
  {Robust {Excitons} and {Trions} in {Monolayer} {MoTe2}}},\ }\href {\doibase
  10.1021/acsnano.5b02665} {\bibfield  {journal} {\bibinfo  {journal} {ACS
  Nano}\ }\textbf {\bibinfo {volume} {9}},\ \bibinfo {pages} {6603} (\bibinfo
  {year} {2015}{\natexlab{a}})}\BibitemShut {NoStop}%
\bibitem [{\citenamefont {Plechinger}\ \emph {et~al.}(2015)\citenamefont
  {Plechinger}, \citenamefont {Nagler}, \citenamefont {Kraus}, \citenamefont
  {Paradiso}, \citenamefont {Strunk}, \citenamefont {Schüller},\ and\
  \citenamefont {Korn}}]{Plechinger_PSS2015}%
  \BibitemOpen
  \bibfield  {author} {\bibinfo {author} {\bibfnamefont {G.}~\bibnamefont
  {Plechinger}}, \bibinfo {author} {\bibfnamefont {P.}~\bibnamefont {Nagler}},
  \bibinfo {author} {\bibfnamefont {J.}~\bibnamefont {Kraus}}, \bibinfo
  {author} {\bibfnamefont {N.}~\bibnamefont {Paradiso}}, \bibinfo {author}
  {\bibfnamefont {C.}~\bibnamefont {Strunk}}, \bibinfo {author} {\bibfnamefont
  {C.}~\bibnamefont {Schüller}}, \ and\ \bibinfo {author} {\bibfnamefont
  {T.}~\bibnamefont {Korn}},\ }\bibfield  {title} {\bibinfo {title} {\emph
  {Identification of excitons, trions and biexcitons in single-layer WS2}},\
  }\href {\doibase https://doi.org/10.1002/pssr.201510224} {\bibfield
  {journal} {\bibinfo  {journal} {physica status solidi (RRL) – Rapid
  Research Letters}\ }\textbf {\bibinfo {volume} {9}},\ \bibinfo {pages} {457}
  (\bibinfo {year} {2015})}\BibitemShut {NoStop}%
\bibitem [{\citenamefont {Singh}\ \emph {et~al.}(2016)\citenamefont {Singh},
  \citenamefont {Moody}, \citenamefont {Tran}, \citenamefont {Scott},
  \citenamefont {Overbeck}, \citenamefont {Bergh\"auser}, \citenamefont
  {Schaibley}, \citenamefont {Seifert}, \citenamefont {Pleskot}, \citenamefont
  {Gabor}, \citenamefont {Yan}, \citenamefont {Mandrus}, \citenamefont
  {Richter}, \citenamefont {Malic}, \citenamefont {Xu},\ and\ \citenamefont
  {Li}}]{Singh2016}%
  \BibitemOpen
  \bibfield  {author} {\bibinfo {author} {\bibfnamefont {A.}~\bibnamefont
  {Singh}}, \bibinfo {author} {\bibfnamefont {G.}~\bibnamefont {Moody}},
  \bibinfo {author} {\bibfnamefont {K.}~\bibnamefont {Tran}}, \bibinfo {author}
  {\bibfnamefont {M.~E.}\ \bibnamefont {Scott}}, \bibinfo {author}
  {\bibfnamefont {V.}~\bibnamefont {Overbeck}}, \bibinfo {author}
  {\bibfnamefont {G.}~\bibnamefont {Bergh\"auser}}, \bibinfo {author}
  {\bibfnamefont {J.}~\bibnamefont {Schaibley}}, \bibinfo {author}
  {\bibfnamefont {E.~J.}\ \bibnamefont {Seifert}}, \bibinfo {author}
  {\bibfnamefont {D.}~\bibnamefont {Pleskot}}, \bibinfo {author} {\bibfnamefont
  {N.~M.}\ \bibnamefont {Gabor}}, \bibinfo {author} {\bibfnamefont
  {J.}~\bibnamefont {Yan}}, \bibinfo {author} {\bibfnamefont {D.~G.}\
  \bibnamefont {Mandrus}}, \bibinfo {author} {\bibfnamefont {M.}~\bibnamefont
  {Richter}}, \bibinfo {author} {\bibfnamefont {E.}~\bibnamefont {Malic}},
  \bibinfo {author} {\bibfnamefont {X.}~\bibnamefont {Xu}}, \ and\ \bibinfo
  {author} {\bibfnamefont {X.}~\bibnamefont {Li}},\ }\bibfield  {title}
  {\bibinfo {title} {\emph {Trion formation dynamics in monolayer transition
  metal dichalcogenides}},\ }\href {\doibase 10.1103/PhysRevB.93.041401}
  {\bibfield  {journal} {\bibinfo  {journal} {Phys. Rev. B}\ }\textbf {\bibinfo
  {volume} {93}},\ \bibinfo {pages} {041401} (\bibinfo {year}
  {2016})}\BibitemShut {NoStop}%
\bibitem [{\citenamefont {Arora}\ \emph {et~al.}(2019)\citenamefont {Arora},
  \citenamefont {Deilmann}, \citenamefont {Reichenauer}, \citenamefont {Kern},
  \citenamefont {Michaelis~de Vasconcellos}, \citenamefont {Rohlfing},\ and\
  \citenamefont {Bratschitsch}}]{Arora2019}%
  \BibitemOpen
  \bibfield  {author} {\bibinfo {author} {\bibfnamefont {A.}~\bibnamefont
  {Arora}}, \bibinfo {author} {\bibfnamefont {T.}~\bibnamefont {Deilmann}},
  \bibinfo {author} {\bibfnamefont {T.}~\bibnamefont {Reichenauer}}, \bibinfo
  {author} {\bibfnamefont {J.}~\bibnamefont {Kern}}, \bibinfo {author}
  {\bibfnamefont {S.}~\bibnamefont {Michaelis~de Vasconcellos}}, \bibinfo
  {author} {\bibfnamefont {M.}~\bibnamefont {Rohlfing}}, \ and\ \bibinfo
  {author} {\bibfnamefont {R.}~\bibnamefont {Bratschitsch}},\ }\bibfield
  {title} {\bibinfo {title} {\emph {Excited-State Trions in Monolayer
  ${\mathrm{WS}}_{2}$}},\ }\href {\doibase 10.1103/PhysRevLett.123.167401}
  {\bibfield  {journal} {\bibinfo  {journal} {Phys. Rev. Lett.}\ }\textbf
  {\bibinfo {volume} {123}},\ \bibinfo {pages} {167401} (\bibinfo {year}
  {2019})}\BibitemShut {NoStop}%
\bibitem [{\citenamefont {Wang}\ \emph {et~al.}(2018)\citenamefont {Wang},
  \citenamefont {Chernikov}, \citenamefont {Glazov}, \citenamefont {Heinz},
  \citenamefont {Marie}, \citenamefont {Amand},\ and\ \citenamefont
  {Urbaszek}}]{WangRev2018}%
  \BibitemOpen
  \bibfield  {author} {\bibinfo {author} {\bibfnamefont {G.}~\bibnamefont
  {Wang}}, \bibinfo {author} {\bibfnamefont {A.}~\bibnamefont {Chernikov}},
  \bibinfo {author} {\bibfnamefont {M.~M.}\ \bibnamefont {Glazov}}, \bibinfo
  {author} {\bibfnamefont {T.~F.}\ \bibnamefont {Heinz}}, \bibinfo {author}
  {\bibfnamefont {X.}~\bibnamefont {Marie}}, \bibinfo {author} {\bibfnamefont
  {T.}~\bibnamefont {Amand}}, \ and\ \bibinfo {author} {\bibfnamefont
  {B.}~\bibnamefont {Urbaszek}},\ }\bibfield  {title} {\bibinfo {title} {\emph
  {Colloquium: Excitons in atomically thin transition metal dichalcogenides}},\
  }\href {\doibase 10.1103/RevModPhys.90.021001} {\bibfield  {journal}
  {\bibinfo  {journal} {Rev. Mod. Phys.}\ }\textbf {\bibinfo {volume} {90}},\
  \bibinfo {pages} {021001} (\bibinfo {year} {2018})}\BibitemShut {NoStop}%
\bibitem [{\citenamefont {Durnev}\ and\ \citenamefont
  {Glazov}(2018)}]{Durnev2018}%
  \BibitemOpen
  \bibfield  {author} {\bibinfo {author} {\bibfnamefont {M.~V.}\ \bibnamefont
  {Durnev}}\ and\ \bibinfo {author} {\bibfnamefont {M.~M.}\ \bibnamefont
  {Glazov}},\ }\bibfield  {title} {\bibinfo {title} {\emph {Excitons and trions
  in two-dimensional semiconductors based on transition metal
  dichalcogenides}},\ }\href {\doibase 10.3367/UFNe.2017.07.038172} {\bibfield
  {journal} {\bibinfo  {journal} {Physics-Uspekhi}\ }\textbf {\bibinfo {volume}
  {61}},\ \bibinfo {pages} {825} (\bibinfo {year} {2018})}\BibitemShut
  {NoStop}%
\bibitem [{\citenamefont {Finkelstein}\ \emph {et~al.}(1996)\citenamefont
  {Finkelstein}, \citenamefont {Shtrikman},\ and\ \citenamefont
  {Bar-Joseph}}]{Finkelstein1996}%
  \BibitemOpen
  \bibfield  {author} {\bibinfo {author} {\bibfnamefont {G.}~\bibnamefont
  {Finkelstein}}, \bibinfo {author} {\bibfnamefont {H.}~\bibnamefont
  {Shtrikman}}, \ and\ \bibinfo {author} {\bibfnamefont {I.}~\bibnamefont
  {Bar-Joseph}},\ }\bibfield  {title} {\bibinfo {title} {\emph {Negatively and
  positively charged excitons in
  $\mathrm{GaAs}/{\mathrm{Al}}_{x}{\mathrm{Ga}}_{1\ensuremath{-}x}\mathrm{As}$
  quantum wells}},\ }\href {\doibase 10.1103/PhysRevB.53.R1709} {\bibfield
  {journal} {\bibinfo  {journal} {Phys. Rev. B}\ }\textbf {\bibinfo {volume}
  {53}},\ \bibinfo {pages} {R1709} (\bibinfo {year} {1996})}\BibitemShut
  {NoStop}%
\bibitem [{\citenamefont {Bracker}\ \emph {et~al.}(2005)\citenamefont
  {Bracker}, \citenamefont {Stinaff}, \citenamefont {Gammon}, \citenamefont
  {Ware}, \citenamefont {Tischler}, \citenamefont {Park}, \citenamefont
  {Gershoni}, \citenamefont {Filinov}, \citenamefont {Bonitz}, \citenamefont
  {Peeters},\ and\ \citenamefont {Riva}}]{Bracker_PRB2005}%
  \BibitemOpen
  \bibfield  {author} {\bibinfo {author} {\bibfnamefont {A.~S.}\ \bibnamefont
  {Bracker}}, \bibinfo {author} {\bibfnamefont {E.~A.}\ \bibnamefont
  {Stinaff}}, \bibinfo {author} {\bibfnamefont {D.}~\bibnamefont {Gammon}},
  \bibinfo {author} {\bibfnamefont {M.~E.}\ \bibnamefont {Ware}}, \bibinfo
  {author} {\bibfnamefont {J.~G.}\ \bibnamefont {Tischler}}, \bibinfo {author}
  {\bibfnamefont {D.}~\bibnamefont {Park}}, \bibinfo {author} {\bibfnamefont
  {D.}~\bibnamefont {Gershoni}}, \bibinfo {author} {\bibfnamefont {A.~V.}\
  \bibnamefont {Filinov}}, \bibinfo {author} {\bibfnamefont {M.}~\bibnamefont
  {Bonitz}}, \bibinfo {author} {\bibfnamefont {F.}~\bibnamefont {Peeters}}, \
  and\ \bibinfo {author} {\bibfnamefont {C.}~\bibnamefont {Riva}},\ }\bibfield
  {title} {\bibinfo {title} {\emph {Binding energies of positive and negative
  trions: From quantum wells to quantum dots}},\ }\href {\doibase
  10.1103/PhysRevB.72.035332} {\bibfield  {journal} {\bibinfo  {journal} {Phys.
  Rev. B}\ }\textbf {\bibinfo {volume} {72}},\ \bibinfo {pages} {035332}
  (\bibinfo {year} {2005})}\BibitemShut {NoStop}%
\bibitem [{\citenamefont {Bar-Joseph}(2005)}]{Joseph2005}%
  \BibitemOpen
  \bibfield  {author} {\bibinfo {author} {\bibfnamefont {I.}~\bibnamefont
  {Bar-Joseph}},\ }\bibfield  {title} {\bibinfo {title} {\emph {Trions in
  {GaAs} quantum wells}},\ }\href {\doibase 10.1088/0268-1242/20/6/R01}
  {\bibfield  {journal} {\bibinfo  {journal} {Semiconductor Science and
  Technology}\ }\textbf {\bibinfo {volume} {20}},\ \bibinfo {pages} {R29}
  (\bibinfo {year} {2005})}\BibitemShut {NoStop}%
\bibitem [{\citenamefont {Kheng}\ \emph {et~al.}(1993)\citenamefont {Kheng},
  \citenamefont {Cox}, \citenamefont {d'~Aubign\'e}, \citenamefont {Bassani},
  \citenamefont {Saminadayar},\ and\ \citenamefont {Tatarenko}}]{Kheng1993}%
  \BibitemOpen
  \bibfield  {author} {\bibinfo {author} {\bibfnamefont {K.}~\bibnamefont
  {Kheng}}, \bibinfo {author} {\bibfnamefont {R.~T.}\ \bibnamefont {Cox}},
  \bibinfo {author} {\bibfnamefont {M.~Y.}\ \bibnamefont {d'~Aubign\'e}},
  \bibinfo {author} {\bibfnamefont {F.}~\bibnamefont {Bassani}}, \bibinfo
  {author} {\bibfnamefont {K.}~\bibnamefont {Saminadayar}}, \ and\ \bibinfo
  {author} {\bibfnamefont {S.}~\bibnamefont {Tatarenko}},\ }\bibfield  {title}
  {\bibinfo {title} {\emph {Observation of negatively charged excitons
  ${\mathit{X}}^{\mathrm{\ensuremath{-}}}$ in semiconductor quantum wells}},\
  }\href {\doibase 10.1103/PhysRevLett.71.1752} {\bibfield  {journal} {\bibinfo
   {journal} {Phys. Rev. Lett.}\ }\textbf {\bibinfo {volume} {71}},\ \bibinfo
  {pages} {1752} (\bibinfo {year} {1993})}\BibitemShut {NoStop}%
\bibitem [{\citenamefont {Huard}\ \emph {et~al.}(2000)\citenamefont {Huard},
  \citenamefont {Cox}, \citenamefont {Saminadayar}, \citenamefont {Arnoult},\
  and\ \citenamefont {Tatarenko}}]{Huard_PRL2000}%
  \BibitemOpen
  \bibfield  {author} {\bibinfo {author} {\bibfnamefont {V.}~\bibnamefont
  {Huard}}, \bibinfo {author} {\bibfnamefont {R.~T.}\ \bibnamefont {Cox}},
  \bibinfo {author} {\bibfnamefont {K.}~\bibnamefont {Saminadayar}}, \bibinfo
  {author} {\bibfnamefont {A.}~\bibnamefont {Arnoult}}, \ and\ \bibinfo
  {author} {\bibfnamefont {S.}~\bibnamefont {Tatarenko}},\ }\bibfield  {title}
  {\bibinfo {title} {\emph {Bound States in Optical Absorption of Semiconductor
  Quantum Wells Containing a Two-Dimensional Electron Gas}},\ }\href {\doibase
  10.1103/PhysRevLett.84.187} {\bibfield  {journal} {\bibinfo  {journal} {Phys.
  Rev. Lett.}\ }\textbf {\bibinfo {volume} {84}},\ \bibinfo {pages} {187}
  (\bibinfo {year} {2000})}\BibitemShut {NoStop}%
\bibitem [{\citenamefont {Portella-Oberli}\ \emph {et~al.}(2004)\citenamefont
  {Portella-Oberli}, \citenamefont {Ciulin}, \citenamefont {Berney},
  \citenamefont {Deveaud}, \citenamefont {Kutrowski},\ and\ \citenamefont
  {Wojtowicz}}]{Portella2004}%
  \BibitemOpen
  \bibfield  {author} {\bibinfo {author} {\bibfnamefont {M.~T.}\ \bibnamefont
  {Portella-Oberli}}, \bibinfo {author} {\bibfnamefont {V.}~\bibnamefont
  {Ciulin}}, \bibinfo {author} {\bibfnamefont {J.~H.}\ \bibnamefont {Berney}},
  \bibinfo {author} {\bibfnamefont {B.}~\bibnamefont {Deveaud}}, \bibinfo
  {author} {\bibfnamefont {M.}~\bibnamefont {Kutrowski}}, \ and\ \bibinfo
  {author} {\bibfnamefont {T.}~\bibnamefont {Wojtowicz}},\ }\bibfield  {title}
  {\bibinfo {title} {\emph {Interacting many-body systems in quantum wells:
  Evidence for exciton-trion-electron correlations}},\ }\href {\doibase
  10.1103/PhysRevB.69.235311} {\bibfield  {journal} {\bibinfo  {journal} {Phys.
  Rev. B}\ }\textbf {\bibinfo {volume} {69}},\ \bibinfo {pages} {235311}
  (\bibinfo {year} {2004})}\BibitemShut {NoStop}%
\bibitem [{\citenamefont {Moody}\ \emph {et~al.}(2014)\citenamefont {Moody},
  \citenamefont {Akimov}, \citenamefont {Li}, \citenamefont {Singh},
  \citenamefont {Yakovlev}, \citenamefont {Karczewski}, \citenamefont {Wiater},
  \citenamefont {Wojtowicz}, \citenamefont {Bayer},\ and\ \citenamefont
  {Cundiff}}]{Moody2014}%
  \BibitemOpen
  \bibfield  {author} {\bibinfo {author} {\bibfnamefont {G.}~\bibnamefont
  {Moody}}, \bibinfo {author} {\bibfnamefont {I.~A.}\ \bibnamefont {Akimov}},
  \bibinfo {author} {\bibfnamefont {H.}~\bibnamefont {Li}}, \bibinfo {author}
  {\bibfnamefont {R.}~\bibnamefont {Singh}}, \bibinfo {author} {\bibfnamefont
  {D.~R.}\ \bibnamefont {Yakovlev}}, \bibinfo {author} {\bibfnamefont
  {G.}~\bibnamefont {Karczewski}}, \bibinfo {author} {\bibfnamefont
  {M.}~\bibnamefont {Wiater}}, \bibinfo {author} {\bibfnamefont
  {T.}~\bibnamefont {Wojtowicz}}, \bibinfo {author} {\bibfnamefont
  {M.}~\bibnamefont {Bayer}}, \ and\ \bibinfo {author} {\bibfnamefont {S.~T.}\
  \bibnamefont {Cundiff}},\ }\bibfield  {title} {\bibinfo {title} {\emph
  {Coherent Coupling of Excitons and Trions in a Photoexcited CdTe/CdMgTe
  Quantum Well}},\ }\href {\doibase 10.1103/PhysRevLett.112.097401} {\bibfield
  {journal} {\bibinfo  {journal} {Phys. Rev. Lett.}\ }\textbf {\bibinfo
  {volume} {112}},\ \bibinfo {pages} {097401} (\bibinfo {year}
  {2014})}\BibitemShut {NoStop}%
\bibitem [{\citenamefont {Astakhov}\ \emph {et~al.}(2002)\citenamefont
  {Astakhov}, \citenamefont {Yakovlev}, \citenamefont {Kochereshko},
  \citenamefont {Ossau}, \citenamefont {Faschinger}, \citenamefont {Puls},
  \citenamefont {Henneberger}, \citenamefont {Crooker}, \citenamefont
  {McCulloch}, \citenamefont {Wolverson}, \citenamefont {Gippius},\ and\
  \citenamefont {Waag}}]{Astakhov2002}%
  \BibitemOpen
  \bibfield  {author} {\bibinfo {author} {\bibfnamefont {G.~V.}\ \bibnamefont
  {Astakhov}}, \bibinfo {author} {\bibfnamefont {D.~R.}\ \bibnamefont
  {Yakovlev}}, \bibinfo {author} {\bibfnamefont {V.~P.}\ \bibnamefont
  {Kochereshko}}, \bibinfo {author} {\bibfnamefont {W.}~\bibnamefont {Ossau}},
  \bibinfo {author} {\bibfnamefont {W.}~\bibnamefont {Faschinger}}, \bibinfo
  {author} {\bibfnamefont {J.}~\bibnamefont {Puls}}, \bibinfo {author}
  {\bibfnamefont {F.}~\bibnamefont {Henneberger}}, \bibinfo {author}
  {\bibfnamefont {S.~A.}\ \bibnamefont {Crooker}}, \bibinfo {author}
  {\bibfnamefont {Q.}~\bibnamefont {McCulloch}}, \bibinfo {author}
  {\bibfnamefont {D.}~\bibnamefont {Wolverson}}, \bibinfo {author}
  {\bibfnamefont {N.~A.}\ \bibnamefont {Gippius}}, \ and\ \bibinfo {author}
  {\bibfnamefont {A.}~\bibnamefont {Waag}},\ }\bibfield  {title} {\bibinfo
  {title} {\emph {Binding energy of charged excitons in ZnSe-based quantum
  wells}},\ }\href {\doibase 10.1103/PhysRevB.65.165335} {\bibfield  {journal}
  {\bibinfo  {journal} {Phys. Rev. B}\ }\textbf {\bibinfo {volume} {65}},\
  \bibinfo {pages} {165335} (\bibinfo {year} {2002})}\BibitemShut {NoStop}%
\bibitem [{\citenamefont {Mueller}\ and\ \citenamefont
  {Malic}(2018)}]{Mueller2018}%
  \BibitemOpen
  \bibfield  {author} {\bibinfo {author} {\bibfnamefont {T.}~\bibnamefont
  {Mueller}}\ and\ \bibinfo {author} {\bibfnamefont {E.}~\bibnamefont
  {Malic}},\ }\bibfield  {title} {\bibinfo {title} {\emph {Exciton physics and
  device application of two-dimensional transition metal dichalcogenide
  semiconductors}},\ }\href {\doibase 10.1038/s41699-018-0074-2} {\bibfield
  {journal} {\bibinfo  {journal} {npj 2D Materials and Applications}\ }\textbf
  {\bibinfo {volume} {2}},\ \bibinfo {pages} {29} (\bibinfo {year}
  {2018})}\BibitemShut {NoStop}%
\bibitem [{\citenamefont {Emmanuele}\ \emph {et~al.}(2020)\citenamefont
  {Emmanuele}, \citenamefont {Sich}, \citenamefont {Kyriienko}, \citenamefont
  {Shahnazaryan}, \citenamefont {Withers}, \citenamefont {Catanzaro},
  \citenamefont {Walker}, \citenamefont {Benimetskiy}, \citenamefont
  {Skolnick}, \citenamefont {Tartakovskii}, \citenamefont {Shelykh},\ and\
  \citenamefont {Krizhanovskii}}]{Emmanuele2020}%
  \BibitemOpen
  \bibfield  {author} {\bibinfo {author} {\bibfnamefont {R.~P.~A.}\
  \bibnamefont {Emmanuele}}, \bibinfo {author} {\bibfnamefont {M.}~\bibnamefont
  {Sich}}, \bibinfo {author} {\bibfnamefont {O.}~\bibnamefont {Kyriienko}},
  \bibinfo {author} {\bibfnamefont {V.}~\bibnamefont {Shahnazaryan}}, \bibinfo
  {author} {\bibfnamefont {F.}~\bibnamefont {Withers}}, \bibinfo {author}
  {\bibfnamefont {A.}~\bibnamefont {Catanzaro}}, \bibinfo {author}
  {\bibfnamefont {P.~M.}\ \bibnamefont {Walker}}, \bibinfo {author}
  {\bibfnamefont {F.~A.}\ \bibnamefont {Benimetskiy}}, \bibinfo {author}
  {\bibfnamefont {M.~S.}\ \bibnamefont {Skolnick}}, \bibinfo {author}
  {\bibfnamefont {A.~I.}\ \bibnamefont {Tartakovskii}}, \bibinfo {author}
  {\bibfnamefont {I.~A.}\ \bibnamefont {Shelykh}}, \ and\ \bibinfo {author}
  {\bibfnamefont {D.~N.}\ \bibnamefont {Krizhanovskii}},\ }\bibfield  {title}
  {\bibinfo {title} {\emph {Highly nonlinear trion-polaritons in a monolayer
  semiconductor}},\ }\href {\doibase 10.1038/s41467-020-17340-z} {\bibfield
  {journal} {\bibinfo  {journal} {Nature Communications}\ }\textbf {\bibinfo
  {volume} {11}},\ \bibinfo {pages} {3589} (\bibinfo {year}
  {2020})}\BibitemShut {NoStop}%
\bibitem [{\citenamefont {Kwon}\ and\ \citenamefont
  {Tabakin}(1978)}]{Kwon1978}%
  \BibitemOpen
  \bibfield  {author} {\bibinfo {author} {\bibfnamefont {Y.~R.}\ \bibnamefont
  {Kwon}}\ and\ \bibinfo {author} {\bibfnamefont {F.}~\bibnamefont {Tabakin}},\
  }\bibfield  {title} {\bibinfo {title} {\emph {Hadronic atoms in momentum
  space}},\ }\href {\doibase 10.1103/PhysRevC.18.932} {\bibfield  {journal}
  {\bibinfo  {journal} {Phys. Rev. C}\ }\textbf {\bibinfo {volume} {18}},\
  \bibinfo {pages} {932} (\bibinfo {year} {1978})}\BibitemShut {NoStop}%
\bibitem [{\citenamefont {Landau}(1983)}]{Landau1983}%
  \BibitemOpen
  \bibfield  {author} {\bibinfo {author} {\bibfnamefont {R.~H.}\ \bibnamefont
  {Landau}},\ }\bibfield  {title} {\bibinfo {title} {\emph {Coupled bound and
  continuum eigenstates in momentum space}},\ }\href {\doibase
  10.1103/PhysRevC.27.2191} {\bibfield  {journal} {\bibinfo  {journal} {Phys.
  Rev. C}\ }\textbf {\bibinfo {volume} {27}},\ \bibinfo {pages} {2191}
  (\bibinfo {year} {1983})}\BibitemShut {NoStop}%
\bibitem [{\citenamefont {Norbury}\ \emph {et~al.}(1994)\citenamefont
  {Norbury}, \citenamefont {Maung},\ and\ \citenamefont
  {Kahana}}]{Norbury1994}%
  \BibitemOpen
  \bibfield  {author} {\bibinfo {author} {\bibfnamefont {J.~W.}\ \bibnamefont
  {Norbury}}, \bibinfo {author} {\bibfnamefont {K.~M.}\ \bibnamefont {Maung}},
  \ and\ \bibinfo {author} {\bibfnamefont {D.~E.}\ \bibnamefont {Kahana}},\
  }\bibfield  {title} {\bibinfo {title} {\emph {Numerical tests of the Land\'e
  subtraction method for the Coulomb potential in momentum space}},\ }\href
  {\doibase 10.1103/PhysRevA.50.2075} {\bibfield  {journal} {\bibinfo
  {journal} {Phys. Rev. A}\ }\textbf {\bibinfo {volume} {50}},\ \bibinfo
  {pages} {2075} (\bibinfo {year} {1994})}\BibitemShut {NoStop}%
\bibitem [{\citenamefont {Ivanov}\ and\ \citenamefont
  {Mitroy}(2001)}]{Ivanov2001}%
  \BibitemOpen
  \bibfield  {author} {\bibinfo {author} {\bibfnamefont {I.}~\bibnamefont
  {Ivanov}}\ and\ \bibinfo {author} {\bibfnamefont {J.}~\bibnamefont
  {Mitroy}},\ }\bibfield  {title} {\bibinfo {title} {\emph {Treatment of the
  Coulomb singularity in momentum space calculations}},\ }\href {\doibase
  https://doi.org/10.1016/S0010-4655(00)00210-1} {\bibfield  {journal}
  {\bibinfo  {journal} {Computer Physics Communications}\ }\textbf {\bibinfo
  {volume} {134}},\ \bibinfo {pages} {317} (\bibinfo {year}
  {2001})}\BibitemShut {NoStop}%
\bibitem [{\citenamefont {Laird}\ \emph {et~al.}(2022)\citenamefont {Laird},
  \citenamefont {Marchetti}, \citenamefont {Efimkin}, \citenamefont {Parish},\
  and\ \citenamefont {Levinsen}}]{Laird2022}%
  \BibitemOpen
  \bibfield  {author} {\bibinfo {author} {\bibfnamefont {E.}~\bibnamefont
  {Laird}}, \bibinfo {author} {\bibfnamefont {F.~M.}\ \bibnamefont
  {Marchetti}}, \bibinfo {author} {\bibfnamefont {D.~K.}\ \bibnamefont
  {Efimkin}}, \bibinfo {author} {\bibfnamefont {M.~M.}\ \bibnamefont {Parish}},
  \ and\ \bibinfo {author} {\bibfnamefont {J.}~\bibnamefont {Levinsen}},\
  }\bibfield  {title} {\bibinfo {title} {\emph {Rydberg exciton-polaritons in a
  magnetic field}},\ }\href {\doibase 10.1103/PhysRevB.106.125407} {\bibfield
  {journal} {\bibinfo  {journal} {Phys. Rev. B}\ }\textbf {\bibinfo {volume}
  {106}},\ \bibinfo {pages} {125407} (\bibinfo {year} {2022})}\BibitemShut
  {NoStop}%
\bibitem [{\citenamefont {de~la Fuente~Pico}\ \emph {et~al.}(2025)\citenamefont
  {de~la Fuente~Pico}, \citenamefont {Levinsen}, \citenamefont {Laird},
  \citenamefont {Parish},\ and\ \citenamefont
  {Marchetti}}]{deLaFuentePico2025}%
  \BibitemOpen
  \bibfield  {author} {\bibinfo {author} {\bibfnamefont {D.}~\bibnamefont
  {de~la Fuente~Pico}}, \bibinfo {author} {\bibfnamefont {J.}~\bibnamefont
  {Levinsen}}, \bibinfo {author} {\bibfnamefont {E.}~\bibnamefont {Laird}},
  \bibinfo {author} {\bibfnamefont {M.~M.}\ \bibnamefont {Parish}}, \ and\
  \bibinfo {author} {\bibfnamefont {F.~M.}\ \bibnamefont {Marchetti}},\
  }\bibfield  {title} {\bibinfo {title} {\emph {Rydberg excitons and polaritons
  in monolayer transition metal dichalcogenides in a magnetic field}},\ }\href
  {\doibase 10.1103/PhysRevB.111.035432} {\bibfield  {journal} {\bibinfo
  {journal} {Phys. Rev. B}\ }\textbf {\bibinfo {volume} {111}},\ \bibinfo
  {pages} {035432} (\bibinfo {year} {2025})}\BibitemShut {NoStop}%
\bibitem [{\citenamefont {Rytova}(1967)}]{Rytova1967}%
  \BibitemOpen
  \bibfield  {author} {\bibinfo {author} {\bibfnamefont {N.}~\bibnamefont
  {Rytova}},\ }\bibfield  {title} {\bibinfo {title} {\emph {The screened
  potential of a point charge in a thin film}},\ }\href {\doibase
  10.48550/arXiv.1806.00976} {\bibfield  {journal} {\bibinfo  {journal} {Moscow
  University Physics Bulletin}\ }\textbf {\bibinfo {volume} {3}},\ \bibinfo
  {pages} {18} (\bibinfo {year} {1967})}\BibitemShut {NoStop}%
\bibitem [{\citenamefont {Keldysh}(1979)}]{Keldysh1979}%
  \BibitemOpen
  \bibfield  {author} {\bibinfo {author} {\bibfnamefont {L.~V.}\ \bibnamefont
  {Keldysh}},\ }\bibfield  {title} {\bibinfo {title} {\emph {Coulomb
  interaction in thin semiconductor and semimetal films}},\ }\href
  {http://jetpletters.ru/ps/0/article_22207.shtml} {\bibfield  {journal}
  {\bibinfo  {journal} {Sov. JETP Lett.}\ }\textbf {\bibinfo {volume} {29}},\
  \bibinfo {pages} {716} (\bibinfo {year} {1979})}\BibitemShut {NoStop}%
\bibitem [{\citenamefont {Cudazzo}\ \emph {et~al.}(2011)\citenamefont
  {Cudazzo}, \citenamefont {Tokatly},\ and\ \citenamefont
  {Rubio}}]{Cudazzo2011}%
  \BibitemOpen
  \bibfield  {author} {\bibinfo {author} {\bibfnamefont {P.}~\bibnamefont
  {Cudazzo}}, \bibinfo {author} {\bibfnamefont {I.~V.}\ \bibnamefont
  {Tokatly}}, \ and\ \bibinfo {author} {\bibfnamefont {A.}~\bibnamefont
  {Rubio}},\ }\bibfield  {title} {\bibinfo {title} {\emph {Dielectric screening
  in two-dimensional insulators: Implications for excitonic and impurity states
  in graphane}},\ }\href {\doibase 10.1103/PhysRevB.84.085406} {\bibfield
  {journal} {\bibinfo  {journal} {Phys. Rev. B}\ }\textbf {\bibinfo {volume}
  {84}},\ \bibinfo {pages} {085406} (\bibinfo {year} {2011})}\BibitemShut
  {NoStop}%
\bibitem [{\citenamefont {Stadler}\ \emph {et~al.}(1991)\citenamefont
  {Stadler}, \citenamefont {Gl\"ockle},\ and\ \citenamefont
  {Sauer}}]{Stadler1991}%
  \BibitemOpen
  \bibfield  {author} {\bibinfo {author} {\bibfnamefont {A.}~\bibnamefont
  {Stadler}}, \bibinfo {author} {\bibfnamefont {W.}~\bibnamefont {Gl\"ockle}},
  \ and\ \bibinfo {author} {\bibfnamefont {P.~U.}\ \bibnamefont {Sauer}},\
  }\bibfield  {title} {\bibinfo {title} {\emph {Faddeev equations with
  three-nucleon force in momentum space}},\ }\href {\doibase
  10.1103/PhysRevC.44.2319} {\bibfield  {journal} {\bibinfo  {journal} {Phys.
  Rev. C}\ }\textbf {\bibinfo {volume} {44}},\ \bibinfo {pages} {2319}
  (\bibinfo {year} {1991})}\BibitemShut {NoStop}%
\bibitem [{\citenamefont {Demmel}(1997)}]{Demmel1997}%
  \BibitemOpen
  \bibfield  {author} {\bibinfo {author} {\bibfnamefont {J.~W.}\ \bibnamefont
  {Demmel}},\ }\href@noop {} {\emph {\bibinfo {title} {Applied Numerical Linear
  Algebra}}}\ (\bibinfo  {publisher} {SIAM - Society for Industrial and Applied
  Mathematics},\ \bibinfo {year} {1997})\BibitemShut {NoStop}%
\bibitem [{\citenamefont {Hadizadeh}\ and\ \citenamefont
  {Bayegan}(2007)}]{Hadizadeh2007}%
  \BibitemOpen
  \bibfield  {author} {\bibinfo {author} {\bibfnamefont {M.}~\bibnamefont
  {Hadizadeh}}\ and\ \bibinfo {author} {\bibfnamefont {S.}~\bibnamefont
  {Bayegan}},\ }\bibfield  {title} {\bibinfo {title} {\emph {Four-Body Bound
  State Formulation in Three-Dimensional Approach}},\ }\href {\doibase
  10.1007/s00601-006-0169-8} {\bibfield  {journal} {\bibinfo  {journal}
  {Few-Body Systems}\ ,\ \bibinfo {pages} {171}} (\bibinfo {year}
  {2007})}\BibitemShut {NoStop}%
\bibitem [{\citenamefont {Hadizadeh}\ \emph {et~al.}(2012)\citenamefont
  {Hadizadeh}, \citenamefont {Yamashita}, \citenamefont {Tomio}, \citenamefont
  {Delfino},\ and\ \citenamefont {Frederico}}]{Hadizadeh2012}%
  \BibitemOpen
  \bibfield  {author} {\bibinfo {author} {\bibfnamefont {M.~R.}\ \bibnamefont
  {Hadizadeh}}, \bibinfo {author} {\bibfnamefont {M.~T.}\ \bibnamefont
  {Yamashita}}, \bibinfo {author} {\bibfnamefont {L.}~\bibnamefont {Tomio}},
  \bibinfo {author} {\bibfnamefont {A.}~\bibnamefont {Delfino}}, \ and\
  \bibinfo {author} {\bibfnamefont {T.}~\bibnamefont {Frederico}},\ }\bibfield
  {title} {\bibinfo {title} {\emph {Binding and structure of tetramers in the
  scaling limit}},\ }\href {\doibase 10.1103/PhysRevA.85.023610} {\bibfield
  {journal} {\bibinfo  {journal} {Physical Review A}\ }\textbf {\bibinfo
  {volume} {85}},\ \bibinfo {pages} {023610} (\bibinfo {year}
  {2012})}\BibitemShut {NoStop}%
\bibitem [{\citenamefont {Sergeev}\ and\ \citenamefont
  {Suris}(2001{\natexlab{a}})}]{Sergeev2001}%
  \BibitemOpen
  \bibfield  {author} {\bibinfo {author} {\bibfnamefont {R.~A.}\ \bibnamefont
  {Sergeev}}\ and\ \bibinfo {author} {\bibfnamefont {R.~A.}\ \bibnamefont
  {Suris}},\ }\bibfield  {title} {\bibinfo {title} {\emph {Ground-state energy
  of {X}- and {X}+ trions in a two-dimensional quantum well at an arbitrary
  mass ratio}},\ }\href {\doibase 10.1134/1.1366005} {\bibfield  {journal}
  {\bibinfo  {journal} {Physics of the Solid State}\ }\textbf {\bibinfo
  {volume} {43}},\ \bibinfo {pages} {746} (\bibinfo {year}
  {2001}{\natexlab{a}})}\BibitemShut {NoStop}%
\bibitem [{\citenamefont {Courtade}\ \emph {et~al.}(2017)\citenamefont
  {Courtade}, \citenamefont {Semina}, \citenamefont {Manca}, \citenamefont
  {Glazov}, \citenamefont {Robert}, \citenamefont {Cadiz}, \citenamefont
  {Wang}, \citenamefont {Taniguchi}, \citenamefont {Watanabe}, \citenamefont
  {Pierre}, \citenamefont {Escoffier}, \citenamefont {Ivchenko}, \citenamefont
  {Renucci}, \citenamefont {Marie}, \citenamefont {Amand},\ and\ \citenamefont
  {Urbaszek}}]{Courtade2017}%
  \BibitemOpen
  \bibfield  {author} {\bibinfo {author} {\bibfnamefont {E.}~\bibnamefont
  {Courtade}}, \bibinfo {author} {\bibfnamefont {M.}~\bibnamefont {Semina}},
  \bibinfo {author} {\bibfnamefont {M.}~\bibnamefont {Manca}}, \bibinfo
  {author} {\bibfnamefont {M.~M.}\ \bibnamefont {Glazov}}, \bibinfo {author}
  {\bibfnamefont {C.}~\bibnamefont {Robert}}, \bibinfo {author} {\bibfnamefont
  {F.}~\bibnamefont {Cadiz}}, \bibinfo {author} {\bibfnamefont
  {G.}~\bibnamefont {Wang}}, \bibinfo {author} {\bibfnamefont {T.}~\bibnamefont
  {Taniguchi}}, \bibinfo {author} {\bibfnamefont {K.}~\bibnamefont {Watanabe}},
  \bibinfo {author} {\bibfnamefont {M.}~\bibnamefont {Pierre}}, \bibinfo
  {author} {\bibfnamefont {W.}~\bibnamefont {Escoffier}}, \bibinfo {author}
  {\bibfnamefont {E.~L.}\ \bibnamefont {Ivchenko}}, \bibinfo {author}
  {\bibfnamefont {P.}~\bibnamefont {Renucci}}, \bibinfo {author} {\bibfnamefont
  {X.}~\bibnamefont {Marie}}, \bibinfo {author} {\bibfnamefont
  {T.}~\bibnamefont {Amand}}, \ and\ \bibinfo {author} {\bibfnamefont
  {B.}~\bibnamefont {Urbaszek}},\ }\bibfield  {title} {\bibinfo {title} {\emph
  {Charged excitons in monolayer ${\mathrm{WSe}}_{2}$: Experiment and
  theory}},\ }\href {\doibase 10.1103/PhysRevB.96.085302} {\bibfield  {journal}
  {\bibinfo  {journal} {Phys. Rev. B}\ }\textbf {\bibinfo {volume} {96}},\
  \bibinfo {pages} {085302} (\bibinfo {year} {2017})}\BibitemShut {NoStop}%
\bibitem [{\citenamefont {Szyniszewski}\ \emph {et~al.}(2017)\citenamefont
  {Szyniszewski}, \citenamefont {Mostaani}, \citenamefont {Drummond},\ and\
  \citenamefont {Fal'ko}}]{Szyniszewski2017}%
  \BibitemOpen
  \bibfield  {author} {\bibinfo {author} {\bibfnamefont {M.}~\bibnamefont
  {Szyniszewski}}, \bibinfo {author} {\bibfnamefont {E.}~\bibnamefont
  {Mostaani}}, \bibinfo {author} {\bibfnamefont {N.~D.}\ \bibnamefont
  {Drummond}}, \ and\ \bibinfo {author} {\bibfnamefont {V.~I.}\ \bibnamefont
  {Fal'ko}},\ }\bibfield  {title} {\bibinfo {title} {\emph {Binding energies of
  trions and biexcitons in two-dimensional semiconductors from diffusion
  quantum Monte Carlo calculations}},\ }\href {\doibase
  10.1103/PhysRevB.95.081301} {\bibfield  {journal} {\bibinfo  {journal} {Phys.
  Rev. B}\ }\textbf {\bibinfo {volume} {95}},\ \bibinfo {pages} {081301}
  (\bibinfo {year} {2017})}\BibitemShut {NoStop}%
\bibitem [{\citenamefont {Fey}\ \emph {et~al.}(2020)\citenamefont {Fey},
  \citenamefont {Schmelcher}, \citenamefont {Imamoglu},\ and\ \citenamefont
  {Schmidt}}]{Fey2020}%
  \BibitemOpen
  \bibfield  {author} {\bibinfo {author} {\bibfnamefont {C.}~\bibnamefont
  {Fey}}, \bibinfo {author} {\bibfnamefont {P.}~\bibnamefont {Schmelcher}},
  \bibinfo {author} {\bibfnamefont {A.}~\bibnamefont {Imamoglu}}, \ and\
  \bibinfo {author} {\bibfnamefont {R.}~\bibnamefont {Schmidt}},\ }\bibfield
  {title} {\bibinfo {title} {\emph {Theory of exciton-electron scattering in
  atomically thin semiconductors}},\ }\href {\doibase
  10.1103/PhysRevB.101.195417} {\bibfield  {journal} {\bibinfo  {journal}
  {Phys. Rev. B}\ }\textbf {\bibinfo {volume} {101}},\ \bibinfo {pages}
  {195417} (\bibinfo {year} {2020})}\BibitemShut {NoStop}%
\bibitem [{\citenamefont {Berkelbach}\ \emph {et~al.}(2013)\citenamefont
  {Berkelbach}, \citenamefont {Hybertsen},\ and\ \citenamefont
  {Reichman}}]{Berkelbach2013}%
  \BibitemOpen
  \bibfield  {author} {\bibinfo {author} {\bibfnamefont {T.~C.}\ \bibnamefont
  {Berkelbach}}, \bibinfo {author} {\bibfnamefont {M.~S.}\ \bibnamefont
  {Hybertsen}}, \ and\ \bibinfo {author} {\bibfnamefont {D.~R.}\ \bibnamefont
  {Reichman}},\ }\bibfield  {title} {\bibinfo {title} {\emph {Theory of neutral
  and charged excitons in monolayer transition metal dichalcogenides}},\ }\href
  {\doibase 10.1103/PhysRevB.88.045318} {\bibfield  {journal} {\bibinfo
  {journal} {Phys. Rev. B}\ }\textbf {\bibinfo {volume} {88}},\ \bibinfo
  {pages} {045318} (\bibinfo {year} {2013})}\BibitemShut {NoStop}%
\bibitem [{\citenamefont {Ramasubramaniam}(2012)}]{Ramasubramaniam_PRB2012}%
  \BibitemOpen
  \bibfield  {author} {\bibinfo {author} {\bibfnamefont {A.}~\bibnamefont
  {Ramasubramaniam}},\ }\bibfield  {title} {\bibinfo {title} {\emph {Large
  excitonic effects in monolayers of molybdenum and tungsten
  dichalcogenides}},\ }\href {\doibase 10.1103/PhysRevB.86.115409} {\bibfield
  {journal} {\bibinfo  {journal} {Phys. Rev. B}\ }\textbf {\bibinfo {volume}
  {86}},\ \bibinfo {pages} {115409} (\bibinfo {year} {2012})}\BibitemShut
  {NoStop}%
\bibitem [{\citenamefont {Komsa}\ and\ \citenamefont
  {Krasheninnikov}(2012)}]{Komsa_PRB2012}%
  \BibitemOpen
  \bibfield  {author} {\bibinfo {author} {\bibfnamefont {H.-P.}\ \bibnamefont
  {Komsa}}\ and\ \bibinfo {author} {\bibfnamefont {A.~V.}\ \bibnamefont
  {Krasheninnikov}},\ }\bibfield  {title} {\bibinfo {title} {\emph {Effects of
  confinement and environment on the electronic structure and exciton binding
  energy of MoS${}_{2}$ from first principles}},\ }\href {\doibase
  10.1103/PhysRevB.86.241201} {\bibfield  {journal} {\bibinfo  {journal} {Phys.
  Rev. B}\ }\textbf {\bibinfo {volume} {86}},\ \bibinfo {pages} {241201}
  (\bibinfo {year} {2012})}\BibitemShut {NoStop}%
\bibitem [{\citenamefont {Shi}\ \emph {et~al.}(2013)\citenamefont {Shi},
  \citenamefont {Pan}, \citenamefont {Zhang},\ and\ \citenamefont
  {Yakobson}}]{Shi_PRB2013}%
  \BibitemOpen
  \bibfield  {author} {\bibinfo {author} {\bibfnamefont {H.}~\bibnamefont
  {Shi}}, \bibinfo {author} {\bibfnamefont {H.}~\bibnamefont {Pan}}, \bibinfo
  {author} {\bibfnamefont {Y.-W.}\ \bibnamefont {Zhang}}, \ and\ \bibinfo
  {author} {\bibfnamefont {B.~I.}\ \bibnamefont {Yakobson}},\ }\bibfield
  {title} {\bibinfo {title} {\emph {Quasiparticle band structures and optical
  properties of strained monolayer MoS${}_{2}$ and WS${}_{2}$}},\ }\href
  {\doibase 10.1103/PhysRevB.87.155304} {\bibfield  {journal} {\bibinfo
  {journal} {Phys. Rev. B}\ }\textbf {\bibinfo {volume} {87}},\ \bibinfo
  {pages} {155304} (\bibinfo {year} {2013})}\BibitemShut {NoStop}%
\bibitem [{\citenamefont {Zipfel}\ \emph {et~al.}(2018)\citenamefont {Zipfel},
  \citenamefont {Holler}, \citenamefont {Mitioglu}, \citenamefont {Ballottin},
  \citenamefont {Nagler}, \citenamefont {Stier}, \citenamefont {Taniguchi},
  \citenamefont {Watanabe}, \citenamefont {Crooker}, \citenamefont
  {Christianen}, \citenamefont {Korn},\ and\ \citenamefont
  {Chernikov}}]{Zipfel-Chernikov-MF_PRB2018}%
  \BibitemOpen
  \bibfield  {author} {\bibinfo {author} {\bibfnamefont {J.}~\bibnamefont
  {Zipfel}}, \bibinfo {author} {\bibfnamefont {J.}~\bibnamefont {Holler}},
  \bibinfo {author} {\bibfnamefont {A.~A.}\ \bibnamefont {Mitioglu}}, \bibinfo
  {author} {\bibfnamefont {M.~V.}\ \bibnamefont {Ballottin}}, \bibinfo {author}
  {\bibfnamefont {P.}~\bibnamefont {Nagler}}, \bibinfo {author} {\bibfnamefont
  {A.~V.}\ \bibnamefont {Stier}}, \bibinfo {author} {\bibfnamefont
  {T.}~\bibnamefont {Taniguchi}}, \bibinfo {author} {\bibfnamefont
  {K.}~\bibnamefont {Watanabe}}, \bibinfo {author} {\bibfnamefont {S.~A.}\
  \bibnamefont {Crooker}}, \bibinfo {author} {\bibfnamefont {P.~C.~M.}\
  \bibnamefont {Christianen}}, \bibinfo {author} {\bibfnamefont
  {T.}~\bibnamefont {Korn}}, \ and\ \bibinfo {author} {\bibfnamefont
  {A.}~\bibnamefont {Chernikov}},\ }\bibfield  {title} {\bibinfo {title} {\emph
  {Spatial extent of the excited exciton states in ${\mathrm{WS}}_{2}$
  monolayers from diamagnetic shifts}},\ }\href {\doibase
  10.1103/PhysRevB.98.075438} {\bibfield  {journal} {\bibinfo  {journal} {Phys.
  Rev. B}\ }\textbf {\bibinfo {volume} {98}},\ \bibinfo {pages} {075438}
  (\bibinfo {year} {2018})}\BibitemShut {NoStop}%
\bibitem [{\citenamefont {Chuang}\ \emph {et~al.}(1991)\citenamefont {Chuang},
  \citenamefont {Schmitt-Rink}, \citenamefont {Miller},\ and\ \citenamefont
  {Chemla}}]{Chuang1991}%
  \BibitemOpen
  \bibfield  {author} {\bibinfo {author} {\bibfnamefont {S.-L.}\ \bibnamefont
  {Chuang}}, \bibinfo {author} {\bibfnamefont {S.}~\bibnamefont
  {Schmitt-Rink}}, \bibinfo {author} {\bibfnamefont {D.~A.~B.}\ \bibnamefont
  {Miller}}, \ and\ \bibinfo {author} {\bibfnamefont {D.~S.}\ \bibnamefont
  {Chemla}},\ }\bibfield  {title} {\bibinfo {title} {\emph {Exciton
  Green's-function approach to optical absorption in a quantum well with an
  applied electric field}},\ }\href {\doibase 10.1103/PhysRevB.43.1500}
  {\bibfield  {journal} {\bibinfo  {journal} {Phys. Rev. B}\ }\textbf {\bibinfo
  {volume} {43}},\ \bibinfo {pages} {1500} (\bibinfo {year}
  {1991})}\BibitemShut {NoStop}%
\bibitem [{\citenamefont {Press}(2007)}]{NumericRec}%
  \BibitemOpen
  \bibfield  {author} {\bibinfo {author} {\bibfnamefont {W.}~\bibnamefont
  {Press}},\ }\href {https://books.google.com.au/books?id=1aAOdzK3FegC} {\emph
  {\bibinfo {title} {Numerical Recipes 3rd Edition: The Art of Scientific
  Computing}}}\ (\bibinfo  {publisher} {Cambridge University Press},\ \bibinfo
  {year} {2007})\BibitemShut {NoStop}%
\bibitem [{\citenamefont {Kezerashvili}(2019)}]{Kezerashvili_FBSReview2019}%
  \BibitemOpen
  \bibfield  {author} {\bibinfo {author} {\bibfnamefont {R.~Y.}\ \bibnamefont
  {Kezerashvili}},\ }\bibfield  {title} {\bibinfo {title} {\emph {Few-Body
  Systems in Condensed Matter Physics}},\ }\href {\doibase
  10.1007/s00601-019-1520-1} {\bibfield  {journal} {\bibinfo  {journal}
  {Few-Body Systems}\ }\textbf {\bibinfo {volume} {60}},\ \bibinfo {pages} {52}
  (\bibinfo {year} {2019})}\BibitemShut {NoStop}%
\bibitem [{\citenamefont {Usukura}\ \emph {et~al.}(1999)\citenamefont
  {Usukura}, \citenamefont {Suzuki},\ and\ \citenamefont
  {Varga}}]{Usukura1999}%
  \BibitemOpen
  \bibfield  {author} {\bibinfo {author} {\bibfnamefont {J.}~\bibnamefont
  {Usukura}}, \bibinfo {author} {\bibfnamefont {Y.}~\bibnamefont {Suzuki}}, \
  and\ \bibinfo {author} {\bibfnamefont {K.}~\bibnamefont {Varga}},\ }\bibfield
   {title} {\bibinfo {title} {\emph {Stability of two- and three-dimensional
  excitonic complexes}},\ }\href {\doibase 10.1103/PhysRevB.59.5652} {\bibfield
   {journal} {\bibinfo  {journal} {Phys. Rev. B}\ }\textbf {\bibinfo {volume}
  {59}},\ \bibinfo {pages} {5652} (\bibinfo {year} {1999})}\BibitemShut
  {NoStop}%
\bibitem [{\citenamefont {Sergeev}\ and\ \citenamefont
  {Suris}(2001{\natexlab{b}})}]{Sergeev_Nanotechnology2001}%
  \BibitemOpen
  \bibfield  {author} {\bibinfo {author} {\bibfnamefont {R.~A.}\ \bibnamefont
  {Sergeev}}\ and\ \bibinfo {author} {\bibfnamefont {R.~A.}\ \bibnamefont
  {Suris}},\ }\bibfield  {title} {\bibinfo {title} {\emph {Singlet and triplet
  states of X+ and X- trions in two-dimensional quantum wells}},\ }\href
  {\doibase 10.1088/0957-4484/12/4/345} {\bibfield  {journal} {\bibinfo
  {journal} {Nanotechnology}\ }\textbf {\bibinfo {volume} {12}},\ \bibinfo
  {pages} {597} (\bibinfo {year} {2001}{\natexlab{b}})}\BibitemShut {NoStop}%
\bibitem [{\citenamefont {Ganchev}\ \emph {et~al.}(2015)\citenamefont
  {Ganchev}, \citenamefont {Drummond}, \citenamefont {Aleiner},\ and\
  \citenamefont {Fal'ko}}]{Ganchev2015}%
  \BibitemOpen
  \bibfield  {author} {\bibinfo {author} {\bibfnamefont {B.}~\bibnamefont
  {Ganchev}}, \bibinfo {author} {\bibfnamefont {N.}~\bibnamefont {Drummond}},
  \bibinfo {author} {\bibfnamefont {I.}~\bibnamefont {Aleiner}}, \ and\
  \bibinfo {author} {\bibfnamefont {V.}~\bibnamefont {Fal'ko}},\ }\bibfield
  {title} {\bibinfo {title} {\emph {Three-Particle Complexes in Two-Dimensional
  Semiconductors}},\ }\href {\doibase 10.1103/PhysRevLett.114.107401}
  {\bibfield  {journal} {\bibinfo  {journal} {Phys. Rev. Lett.}\ }\textbf
  {\bibinfo {volume} {114}},\ \bibinfo {pages} {107401} (\bibinfo {year}
  {2015})}\BibitemShut {NoStop}%
\bibitem [{\citenamefont {Kidd}\ \emph {et~al.}(2016)\citenamefont {Kidd},
  \citenamefont {Zhang},\ and\ \citenamefont {Varga}}]{Kidd_PRB2016}%
  \BibitemOpen
  \bibfield  {author} {\bibinfo {author} {\bibfnamefont {D.~W.}\ \bibnamefont
  {Kidd}}, \bibinfo {author} {\bibfnamefont {D.~K.}\ \bibnamefont {Zhang}}, \
  and\ \bibinfo {author} {\bibfnamefont {K.}~\bibnamefont {Varga}},\ }\bibfield
   {title} {\bibinfo {title} {\emph {Binding energies and structures of
  two-dimensional excitonic complexes in transition metal dichalcogenides}},\
  }\href {\doibase 10.1103/PhysRevB.93.125423} {\bibfield  {journal} {\bibinfo
  {journal} {Phys. Rev. B}\ }\textbf {\bibinfo {volume} {93}},\ \bibinfo
  {pages} {125423} (\bibinfo {year} {2016})}\BibitemShut {NoStop}%
\bibitem [{\citenamefont {Mayers}\ \emph {et~al.}(2015)\citenamefont {Mayers},
  \citenamefont {Berkelbach}, \citenamefont {Hybertsen},\ and\ \citenamefont
  {Reichman}}]{Mayers2015}%
  \BibitemOpen
  \bibfield  {author} {\bibinfo {author} {\bibfnamefont {M.~Z.}\ \bibnamefont
  {Mayers}}, \bibinfo {author} {\bibfnamefont {T.~C.}\ \bibnamefont
  {Berkelbach}}, \bibinfo {author} {\bibfnamefont {M.~S.}\ \bibnamefont
  {Hybertsen}}, \ and\ \bibinfo {author} {\bibfnamefont {D.~R.}\ \bibnamefont
  {Reichman}},\ }\bibfield  {title} {\bibinfo {title} {\emph {Binding energies
  and spatial structures of small carrier complexes in monolayer
  transition-metal dichalcogenides via diffusion Monte Carlo}},\ }\href
  {\doibase 10.1103/PhysRevB.92.161404} {\bibfield  {journal} {\bibinfo
  {journal} {Phys. Rev. B}\ }\textbf {\bibinfo {volume} {92}},\ \bibinfo
  {pages} {161404} (\bibinfo {year} {2015})}\BibitemShut {NoStop}%
\bibitem [{\citenamefont {Spink}\ \emph {et~al.}(2016)\citenamefont {Spink},
  \citenamefont {L\'opez~R\'{\i}os}, \citenamefont {Drummond},\ and\
  \citenamefont {Needs}}]{Spink2016}%
  \BibitemOpen
  \bibfield  {author} {\bibinfo {author} {\bibfnamefont {G.~G.}\ \bibnamefont
  {Spink}}, \bibinfo {author} {\bibfnamefont {P.}~\bibnamefont
  {L\'opez~R\'{\i}os}}, \bibinfo {author} {\bibfnamefont {N.~D.}\ \bibnamefont
  {Drummond}}, \ and\ \bibinfo {author} {\bibfnamefont {R.~J.}\ \bibnamefont
  {Needs}},\ }\bibfield  {title} {\bibinfo {title} {\emph {Trion formation in a
  two-dimensional hole-doped electron gas}},\ }\href {\doibase
  10.1103/PhysRevB.94.041410} {\bibfield  {journal} {\bibinfo  {journal} {Phys.
  Rev. B}\ }\textbf {\bibinfo {volume} {94}},\ \bibinfo {pages} {041410}
  (\bibinfo {year} {2016})}\BibitemShut {NoStop}%
\bibitem [{\citenamefont {Deilmann}\ and\ \citenamefont
  {Thygesen}(2017)}]{Deilmann2017}%
  \BibitemOpen
  \bibfield  {author} {\bibinfo {author} {\bibfnamefont {T.}~\bibnamefont
  {Deilmann}}\ and\ \bibinfo {author} {\bibfnamefont {K.~S.}\ \bibnamefont
  {Thygesen}},\ }\bibfield  {title} {\bibinfo {title} {\emph {Dark excitations
  in monolayer transition metal dichalcogenides}},\ }\href {\doibase
  10.1103/PhysRevB.96.201113} {\bibfield  {journal} {\bibinfo  {journal} {Phys.
  Rev. B}\ }\textbf {\bibinfo {volume} {96}},\ \bibinfo {pages} {201113}
  (\bibinfo {year} {2017})}\BibitemShut {NoStop}%
\bibitem [{\citenamefont {Van~der Donck}\ \emph {et~al.}(2017)\citenamefont
  {Van~der Donck}, \citenamefont {Zarenia},\ and\ \citenamefont
  {Peeters}}]{Donck2017}%
  \BibitemOpen
  \bibfield  {author} {\bibinfo {author} {\bibfnamefont {M.}~\bibnamefont
  {Van~der Donck}}, \bibinfo {author} {\bibfnamefont {M.}~\bibnamefont
  {Zarenia}}, \ and\ \bibinfo {author} {\bibfnamefont {F.~M.}\ \bibnamefont
  {Peeters}},\ }\bibfield  {title} {\bibinfo {title} {\emph {Excitons and
  trions in monolayer transition metal dichalcogenides: A comparative study
  between the multiband model and the quadratic single-band model}},\ }\href
  {\doibase 10.1103/PhysRevB.96.035131} {\bibfield  {journal} {\bibinfo
  {journal} {Phys. Rev. B}\ }\textbf {\bibinfo {volume} {96}},\ \bibinfo
  {pages} {035131} (\bibinfo {year} {2017})}\BibitemShut {NoStop}%
\bibitem [{\citenamefont {Efimkin}\ \emph {et~al.}(2021)\citenamefont
  {Efimkin}, \citenamefont {Laird}, \citenamefont {Levinsen}, \citenamefont
  {Parish},\ and\ \citenamefont {MacDonald}}]{Efimkin2021}%
  \BibitemOpen
  \bibfield  {author} {\bibinfo {author} {\bibfnamefont {D.~K.}\ \bibnamefont
  {Efimkin}}, \bibinfo {author} {\bibfnamefont {E.~K.}\ \bibnamefont {Laird}},
  \bibinfo {author} {\bibfnamefont {J.}~\bibnamefont {Levinsen}}, \bibinfo
  {author} {\bibfnamefont {M.~M.}\ \bibnamefont {Parish}}, \ and\ \bibinfo
  {author} {\bibfnamefont {A.~H.}\ \bibnamefont {MacDonald}},\ }\bibfield
  {title} {\bibinfo {title} {\emph {Electron-exciton interactions in the
  exciton-polaron problem}},\ }\href {\doibase 10.1103/PhysRevB.103.075417}
  {\bibfield  {journal} {\bibinfo  {journal} {Physical Review B}\ }\textbf
  {\bibinfo {volume} {103}},\ \bibinfo {pages} {075417} (\bibinfo {year}
  {2021})}\BibitemShut {NoStop}%
\bibitem [{\citenamefont {Mohseni}\ \emph {et~al.}(2023)\citenamefont
  {Mohseni}, \citenamefont {Hadizadeh}, \citenamefont {Frederico},
  \citenamefont {da~Costa},\ and\ \citenamefont {Chaves}}]{Mohseni2023}%
  \BibitemOpen
  \bibfield  {author} {\bibinfo {author} {\bibfnamefont {K.}~\bibnamefont
  {Mohseni}}, \bibinfo {author} {\bibfnamefont {M.~R.}\ \bibnamefont
  {Hadizadeh}}, \bibinfo {author} {\bibfnamefont {T.}~\bibnamefont
  {Frederico}}, \bibinfo {author} {\bibfnamefont {D.~R.}\ \bibnamefont
  {da~Costa}}, \ and\ \bibinfo {author} {\bibfnamefont {A.~J.}\ \bibnamefont
  {Chaves}},\ }\bibfield  {title} {\bibinfo {title} {\emph {Trion clustering
  structure and binding energy in two-dimensional semiconductor materials:
  Faddeev equations approach}},\ }\href {\doibase 10.1103/PhysRevB.107.165427}
  {\bibfield  {journal} {\bibinfo  {journal} {Phys. Rev. B}\ }\textbf {\bibinfo
  {volume} {107}},\ \bibinfo {pages} {165427} (\bibinfo {year}
  {2023})}\BibitemShut {NoStop}%
\bibitem [{\citenamefont {Filikhin}\ \emph {et~al.}(2018)\citenamefont
  {Filikhin}, \citenamefont {Kezerashvili}, \citenamefont {Tsiklauri},\ and\
  \citenamefont {Vlahovic}}]{Filikhin2018}%
  \BibitemOpen
  \bibfield  {author} {\bibinfo {author} {\bibfnamefont {I.}~\bibnamefont
  {Filikhin}}, \bibinfo {author} {\bibfnamefont {R.~Y.}\ \bibnamefont
  {Kezerashvili}}, \bibinfo {author} {\bibfnamefont {S.~M.}\ \bibnamefont
  {Tsiklauri}}, \ and\ \bibinfo {author} {\bibfnamefont {B.}~\bibnamefont
  {Vlahovic}},\ }\bibfield  {title} {\bibinfo {title} {\emph {Trions in bulk
  and monolayer materials: Faddeev equations and hyperspherical harmonics}},\
  }\href {\doibase 10.1088/1361-6528/aaa94d} {\bibfield  {journal} {\bibinfo
  {journal} {Nanotechnology}\ }\textbf {\bibinfo {volume} {29}},\ \bibinfo
  {pages} {124002} (\bibinfo {year} {2018})}\BibitemShut {NoStop}%
\bibitem [{\citenamefont {Kezerashvili}\ \emph {et~al.}(2024)\citenamefont
  {Kezerashvili}, \citenamefont {Tsiklauri},\ and\ \citenamefont
  {Dublin}}]{Kezerashvili2024}%
  \BibitemOpen
  \bibfield  {author} {\bibinfo {author} {\bibfnamefont {R.~Y.}\ \bibnamefont
  {Kezerashvili}}, \bibinfo {author} {\bibfnamefont {S.~M.}\ \bibnamefont
  {Tsiklauri}}, \ and\ \bibinfo {author} {\bibfnamefont {A.}~\bibnamefont
  {Dublin}},\ }\bibfield  {title} {\bibinfo {title} {\emph {Trions in
  two-dimensional monolayers within the hyperspherical harmonics method:
  Application to transition metal dichalcogenides}},\ }\href {\doibase
  10.1103/PhysRevB.109.085406} {\bibfield  {journal} {\bibinfo  {journal}
  {Phys. Rev. B}\ }\textbf {\bibinfo {volume} {109}},\ \bibinfo {pages}
  {085406} (\bibinfo {year} {2024})}\BibitemShut {NoStop}%
\bibitem [{\citenamefont {Pricoupenko}\ and\ \citenamefont
  {Pedri}(2010)}]{Pricoupenko2010}%
  \BibitemOpen
  \bibfield  {author} {\bibinfo {author} {\bibfnamefont {L.}~\bibnamefont
  {Pricoupenko}}\ and\ \bibinfo {author} {\bibfnamefont {P.}~\bibnamefont
  {Pedri}},\ }\bibfield  {title} {\bibinfo {title} {\emph {Universal
  ($1+2$)-body bound states in planar atomic waveguides}},\ }\href {\doibase
  10.1103/PhysRevA.82.033625} {\bibfield  {journal} {\bibinfo  {journal} {Phys.
  Rev. A}\ }\textbf {\bibinfo {volume} {82}},\ \bibinfo {pages} {033625}
  (\bibinfo {year} {2010})}\BibitemShut {NoStop}%
\bibitem [{\citenamefont {Parish}\ and\ \citenamefont
  {Levinsen}(2013)}]{Parish_PRA2013}%
  \BibitemOpen
  \bibfield  {author} {\bibinfo {author} {\bibfnamefont {M.~M.}\ \bibnamefont
  {Parish}}\ and\ \bibinfo {author} {\bibfnamefont {J.}~\bibnamefont
  {Levinsen}},\ }\bibfield  {title} {\bibinfo {title} {\emph {Highly polarized
  Fermi gases in two dimensions}},\ }\href {\doibase
  10.1103/PhysRevA.87.033616} {\bibfield  {journal} {\bibinfo  {journal} {Phys.
  Rev. A}\ }\textbf {\bibinfo {volume} {87}},\ \bibinfo {pages} {033616}
  (\bibinfo {year} {2013})}\BibitemShut {NoStop}%
\bibitem [{\citenamefont {Ngampruetikorn}\ \emph {et~al.}(2013)\citenamefont
  {Ngampruetikorn}, \citenamefont {Parish},\ and\ \citenamefont
  {Levinsen}}]{Ngampruetikorn2013}%
  \BibitemOpen
  \bibfield  {author} {\bibinfo {author} {\bibfnamefont {V.}~\bibnamefont
  {Ngampruetikorn}}, \bibinfo {author} {\bibfnamefont {M.~M.}\ \bibnamefont
  {Parish}}, \ and\ \bibinfo {author} {\bibfnamefont {J.}~\bibnamefont
  {Levinsen}},\ }\bibfield  {title} {\bibinfo {title} {\emph {Three-body
  problem in a two-dimensional Fermi gas}},\ }\href {\doibase
  10.1209/0295-5075/102/13001} {\bibfield  {journal} {\bibinfo  {journal}
  {{EPL} (Europhysics Letters)}\ }\textbf {\bibinfo {volume} {102}},\ \bibinfo
  {pages} {13001} (\bibinfo {year} {2013})}\BibitemShut {NoStop}%
\bibitem [{\citenamefont {Tiene}\ \emph {et~al.}(2022)\citenamefont {Tiene},
  \citenamefont {Levinsen}, \citenamefont {Keeling}, \citenamefont {Parish},\
  and\ \citenamefont {Marchetti}}]{Tiene2022}%
  \BibitemOpen
  \bibfield  {author} {\bibinfo {author} {\bibfnamefont {A.}~\bibnamefont
  {Tiene}}, \bibinfo {author} {\bibfnamefont {J.}~\bibnamefont {Levinsen}},
  \bibinfo {author} {\bibfnamefont {J.}~\bibnamefont {Keeling}}, \bibinfo
  {author} {\bibfnamefont {M.~M.}\ \bibnamefont {Parish}}, \ and\ \bibinfo
  {author} {\bibfnamefont {F.~M.}\ \bibnamefont {Marchetti}},\ }\bibfield
  {title} {\bibinfo {title} {\emph {Effect of fermion indistinguishability on
  optical absorption of doped two-dimensional semiconductors}},\ }\href
  {\doibase 10.1103/PhysRevB.105.125404} {\bibfield  {journal} {\bibinfo
  {journal} {Phys. Rev. B}\ }\textbf {\bibinfo {volume} {105}},\ \bibinfo
  {pages} {125404} (\bibinfo {year} {2022})}\BibitemShut {NoStop}%
\bibitem [{\citenamefont {Combescot}(2017)}]{CombescotPRX2017}%
  \BibitemOpen
  \bibfield  {author} {\bibinfo {author} {\bibfnamefont {R.}~\bibnamefont
  {Combescot}},\ }\bibfield  {title} {\bibinfo {title} {\emph {Three-Body
  Coulomb Problem}},\ }\href {\doibase 10.1103/PhysRevX.7.041035} {\bibfield
  {journal} {\bibinfo  {journal} {Phys. Rev. X}\ }\textbf {\bibinfo {volume}
  {7}},\ \bibinfo {pages} {041035} (\bibinfo {year} {2017})}\BibitemShut
  {NoStop}%
\bibitem [{\citenamefont {Stébé}\ and\ \citenamefont
  {Ainane}(1989)}]{Stebe1989}%
  \BibitemOpen
  \bibfield  {author} {\bibinfo {author} {\bibfnamefont {B.}~\bibnamefont
  {Stébé}}\ and\ \bibinfo {author} {\bibfnamefont {A.}~\bibnamefont
  {Ainane}},\ }\bibfield  {title} {\bibinfo {title} {\emph {Ground state energy
  and optical absorption of excitonic trions in two dimensional
  semiconductors}},\ }\href {\doibase
  https://doi.org/10.1016/0749-6036(89)90382-0} {\bibfield  {journal} {\bibinfo
   {journal} {Superlattices and Microstructures}\ }\textbf {\bibinfo {volume}
  {5}},\ \bibinfo {pages} {545} (\bibinfo {year} {1989})}\BibitemShut {NoStop}%
\bibitem [{\citenamefont {Rasmussen}\ and\ \citenamefont
  {Thygesen}(2015)}]{Rasmussen_PhysChemC2015}%
  \BibitemOpen
  \bibfield  {author} {\bibinfo {author} {\bibfnamefont {F.~A.}\ \bibnamefont
  {Rasmussen}}\ and\ \bibinfo {author} {\bibfnamefont {K.~S.}\ \bibnamefont
  {Thygesen}},\ }\bibfield  {title} {\bibinfo {title} {\emph {Computational 2D
  Materials Database: Electronic Structure of Transition-Metal Dichalcogenides
  and Oxides}},\ }\href {\doibase 10.1021/acs.jpcc.5b02950} {\bibfield
  {journal} {\bibinfo  {journal} {The Journal of Physical Chemistry C}\
  }\textbf {\bibinfo {volume} {119}},\ \bibinfo {pages} {13169} (\bibinfo
  {year} {2015})}\BibitemShut {NoStop}%
\bibitem [{\citenamefont {Kyl\"anp\"a\"a}\ and\ \citenamefont
  {Komsa}(2015)}]{Kylanpaa2015}%
  \BibitemOpen
  \bibfield  {author} {\bibinfo {author} {\bibfnamefont {I.}~\bibnamefont
  {Kyl\"anp\"a\"a}}\ and\ \bibinfo {author} {\bibfnamefont {H.-P.}\
  \bibnamefont {Komsa}},\ }\bibfield  {title} {\bibinfo {title} {\emph {Binding
  energies of exciton complexes in transition metal dichalcogenide monolayers
  and effect of dielectric environment}},\ }\href {\doibase
  10.1103/PhysRevB.92.205418} {\bibfield  {journal} {\bibinfo  {journal} {Phys.
  Rev. B}\ }\textbf {\bibinfo {volume} {92}},\ \bibinfo {pages} {205418}
  (\bibinfo {year} {2015})}\BibitemShut {NoStop}%
\bibitem [{\citenamefont {Björkman}(2014)}]{Bjorkman2014}%
  \BibitemOpen
  \bibfield  {author} {\bibinfo {author} {\bibfnamefont {T.}~\bibnamefont
  {Björkman}},\ }\bibfield  {title} {\bibinfo {title} {\emph {{Testing several
  recent van der Waals density functionals for layered structures}}},\ }\href
  {\doibase 10.1063/1.4893329} {\bibfield  {journal} {\bibinfo  {journal} {The
  Journal of Chemical Physics}\ }\textbf {\bibinfo {volume} {141}},\ \bibinfo
  {pages} {074708} (\bibinfo {year} {2014})}\BibitemShut {NoStop}%
\bibitem [{\citenamefont {You}\ \emph {et~al.}(2015)\citenamefont {You},
  \citenamefont {Zhang}, \citenamefont {Berkelbach}, \citenamefont {Hybertsen},
  \citenamefont {Reichman},\ and\ \citenamefont
  {Heinz}}]{You-Heinz_NatPhys2015}%
  \BibitemOpen
  \bibfield  {author} {\bibinfo {author} {\bibfnamefont {Y.}~\bibnamefont
  {You}}, \bibinfo {author} {\bibfnamefont {X.-X.}\ \bibnamefont {Zhang}},
  \bibinfo {author} {\bibfnamefont {T.~C.}\ \bibnamefont {Berkelbach}},
  \bibinfo {author} {\bibfnamefont {M.~S.}\ \bibnamefont {Hybertsen}}, \bibinfo
  {author} {\bibfnamefont {D.~R.}\ \bibnamefont {Reichman}}, \ and\ \bibinfo
  {author} {\bibfnamefont {T.~F.}\ \bibnamefont {Heinz}},\ }\bibfield  {title}
  {\bibinfo {title} {\emph {Observation of biexcitons in monolayer WSe2}},\
  }\href {\doibase 10.1038/nphys3324} {\bibfield  {journal} {\bibinfo
  {journal} {Nature Physics}\ }\textbf {\bibinfo {volume} {11}},\ \bibinfo
  {pages} {477} (\bibinfo {year} {2015})}\BibitemShut {NoStop}%
\bibitem [{\citenamefont {Stevens}\ \emph {et~al.}(2018)\citenamefont
  {Stevens}, \citenamefont {Paul}, \citenamefont {Cox}, \citenamefont {Sahoo},
  \citenamefont {Guti{\'e}rrez}, \citenamefont {Turkowski}, \citenamefont
  {Semenov}, \citenamefont {McGill}, \citenamefont {Kapetanakis}, \citenamefont
  {Perakis}, \citenamefont {Hilton},\ and\ \citenamefont
  {Karaiskaj}}]{Stevens_NatComm2018}%
  \BibitemOpen
  \bibfield  {author} {\bibinfo {author} {\bibfnamefont {C.~E.}\ \bibnamefont
  {Stevens}}, \bibinfo {author} {\bibfnamefont {J.}~\bibnamefont {Paul}},
  \bibinfo {author} {\bibfnamefont {T.}~\bibnamefont {Cox}}, \bibinfo {author}
  {\bibfnamefont {P.~K.}\ \bibnamefont {Sahoo}}, \bibinfo {author}
  {\bibfnamefont {H.~R.}\ \bibnamefont {Guti{\'e}rrez}}, \bibinfo {author}
  {\bibfnamefont {V.}~\bibnamefont {Turkowski}}, \bibinfo {author}
  {\bibfnamefont {D.}~\bibnamefont {Semenov}}, \bibinfo {author} {\bibfnamefont
  {S.~A.}\ \bibnamefont {McGill}}, \bibinfo {author} {\bibfnamefont {M.~D.}\
  \bibnamefont {Kapetanakis}}, \bibinfo {author} {\bibfnamefont {I.~E.}\
  \bibnamefont {Perakis}}, \bibinfo {author} {\bibfnamefont {D.~J.}\
  \bibnamefont {Hilton}}, \ and\ \bibinfo {author} {\bibfnamefont
  {D.}~\bibnamefont {Karaiskaj}},\ }\bibfield  {title} {\bibinfo {title} {\emph
  {Biexcitons in monolayer transition metal dichalcogenides tuned by magnetic
  fields}},\ }\href {\doibase 10.1038/s41467-018-05643-1} {\bibfield  {journal}
  {\bibinfo  {journal} {Nature Communications}\ }\textbf {\bibinfo {volume}
  {9}},\ \bibinfo {pages} {3720} (\bibinfo {year} {2018})}\BibitemShut
  {NoStop}%
\bibitem [{\citenamefont {Steinhoff}\ \emph {et~al.}(2018)\citenamefont
  {Steinhoff}, \citenamefont {Florian}, \citenamefont {Singh}, \citenamefont
  {Tran}, \citenamefont {Kolarczik}, \citenamefont {Helmrich}, \citenamefont
  {Achtstein}, \citenamefont {Woggon}, \citenamefont {Owschimikow},
  \citenamefont {Jahnke},\ and\ \citenamefont {Li}}]{Steinhoff-Li_NatPhys2018}%
  \BibitemOpen
  \bibfield  {author} {\bibinfo {author} {\bibfnamefont {A.}~\bibnamefont
  {Steinhoff}}, \bibinfo {author} {\bibfnamefont {M.}~\bibnamefont {Florian}},
  \bibinfo {author} {\bibfnamefont {A.}~\bibnamefont {Singh}}, \bibinfo
  {author} {\bibfnamefont {K.}~\bibnamefont {Tran}}, \bibinfo {author}
  {\bibfnamefont {M.}~\bibnamefont {Kolarczik}}, \bibinfo {author}
  {\bibfnamefont {S.}~\bibnamefont {Helmrich}}, \bibinfo {author}
  {\bibfnamefont {A.~W.}\ \bibnamefont {Achtstein}}, \bibinfo {author}
  {\bibfnamefont {U.}~\bibnamefont {Woggon}}, \bibinfo {author} {\bibfnamefont
  {N.}~\bibnamefont {Owschimikow}}, \bibinfo {author} {\bibfnamefont
  {F.}~\bibnamefont {Jahnke}}, \ and\ \bibinfo {author} {\bibfnamefont
  {X.}~\bibnamefont {Li}},\ }\bibfield  {title} {\bibinfo {title} {\emph
  {Biexciton fine structure in monolayer transition metal dichalcogenides}},\
  }\href {\doibase 10.1038/s41567-018-0282-x} {\bibfield  {journal} {\bibinfo
  {journal} {Nature Physics}\ }\textbf {\bibinfo {volume} {14}},\ \bibinfo
  {pages} {1199} (\bibinfo {year} {2018})}\BibitemShut {NoStop}%
\bibitem [{\citenamefont {Conway}\ \emph {et~al.}(2022)\citenamefont {Conway},
  \citenamefont {Muir}, \citenamefont {Earl}, \citenamefont {Wurdack},
  \citenamefont {Mishra}, \citenamefont {Tollerud},\ and\ \citenamefont
  {Davis}}]{Conway_2DMat2022}%
  \BibitemOpen
  \bibfield  {author} {\bibinfo {author} {\bibfnamefont {M.~A.}\ \bibnamefont
  {Conway}}, \bibinfo {author} {\bibfnamefont {J.~B.}\ \bibnamefont {Muir}},
  \bibinfo {author} {\bibfnamefont {S.~K.}\ \bibnamefont {Earl}}, \bibinfo
  {author} {\bibfnamefont {M.}~\bibnamefont {Wurdack}}, \bibinfo {author}
  {\bibfnamefont {R.}~\bibnamefont {Mishra}}, \bibinfo {author} {\bibfnamefont
  {J.~O.}\ \bibnamefont {Tollerud}}, \ and\ \bibinfo {author} {\bibfnamefont
  {J.~A.}\ \bibnamefont {Davis}},\ }\bibfield  {title} {\bibinfo {title} {\emph
  {Direct measurement of biexcitons in monolayer WS2}},\ }\href {\doibase
  10.1088/2053-1583/ac4779} {\bibfield  {journal} {\bibinfo  {journal} {2D
  Materials}\ }\textbf {\bibinfo {volume} {9}},\ \bibinfo {pages} {021001}
  (\bibinfo {year} {2022})}\BibitemShut {NoStop}%
\bibitem [{\citenamefont {Hao}\ \emph {et~al.}(2017)\citenamefont {Hao},
  \citenamefont {Specht}, \citenamefont {Nagler}, \citenamefont {Xu},
  \citenamefont {Tran}, \citenamefont {Singh}, \citenamefont {Dass},
  \citenamefont {Sch{\"u}ller}, \citenamefont {Korn}, \citenamefont {Richter},
  \citenamefont {Knorr}, \citenamefont {Li},\ and\ \citenamefont
  {Moody}}]{Hao2017}%
  \BibitemOpen
  \bibfield  {author} {\bibinfo {author} {\bibfnamefont {K.}~\bibnamefont
  {Hao}}, \bibinfo {author} {\bibfnamefont {J.~F.}\ \bibnamefont {Specht}},
  \bibinfo {author} {\bibfnamefont {P.}~\bibnamefont {Nagler}}, \bibinfo
  {author} {\bibfnamefont {L.}~\bibnamefont {Xu}}, \bibinfo {author}
  {\bibfnamefont {K.}~\bibnamefont {Tran}}, \bibinfo {author} {\bibfnamefont
  {A.}~\bibnamefont {Singh}}, \bibinfo {author} {\bibfnamefont {C.~K.}\
  \bibnamefont {Dass}}, \bibinfo {author} {\bibfnamefont {C.}~\bibnamefont
  {Sch{\"u}ller}}, \bibinfo {author} {\bibfnamefont {T.}~\bibnamefont {Korn}},
  \bibinfo {author} {\bibfnamefont {M.}~\bibnamefont {Richter}}, \bibinfo
  {author} {\bibfnamefont {A.}~\bibnamefont {Knorr}}, \bibinfo {author}
  {\bibfnamefont {X.}~\bibnamefont {Li}}, \ and\ \bibinfo {author}
  {\bibfnamefont {G.}~\bibnamefont {Moody}},\ }\bibfield  {title} {\bibinfo
  {title} {\emph {Neutral and charged inter-valley biexcitons in monolayer
  MoSe2}},\ }\href {\doibase 10.1038/ncomms15552} {\bibfield  {journal}
  {\bibinfo  {journal} {Nature Communications}\ }\textbf {\bibinfo {volume}
  {8}},\ \bibinfo {pages} {15552} (\bibinfo {year} {2017})}\BibitemShut
  {NoStop}%
\bibitem [{\citenamefont {Barbone}\ \emph {et~al.}(2018)\citenamefont
  {Barbone}, \citenamefont {Montblanch}, \citenamefont {Kara}, \citenamefont
  {Palacios-Berraquero}, \citenamefont {Cadore}, \citenamefont {De~Fazio},
  \citenamefont {Pingault}, \citenamefont {Mostaani}, \citenamefont {Li},
  \citenamefont {Chen}, \citenamefont {Watanabe}, \citenamefont {Taniguchi},
  \citenamefont {Tongay}, \citenamefont {Wang}, \citenamefont {Ferrari},\ and\
  \citenamefont {Atat{\"u}re}}]{Barbone-Atature_NatComm2018}%
  \BibitemOpen
  \bibfield  {author} {\bibinfo {author} {\bibfnamefont {M.}~\bibnamefont
  {Barbone}}, \bibinfo {author} {\bibfnamefont {A.~R.-P.}\ \bibnamefont
  {Montblanch}}, \bibinfo {author} {\bibfnamefont {D.~M.}\ \bibnamefont
  {Kara}}, \bibinfo {author} {\bibfnamefont {C.}~\bibnamefont
  {Palacios-Berraquero}}, \bibinfo {author} {\bibfnamefont {A.~R.}\
  \bibnamefont {Cadore}}, \bibinfo {author} {\bibfnamefont {D.}~\bibnamefont
  {De~Fazio}}, \bibinfo {author} {\bibfnamefont {B.}~\bibnamefont {Pingault}},
  \bibinfo {author} {\bibfnamefont {E.}~\bibnamefont {Mostaani}}, \bibinfo
  {author} {\bibfnamefont {H.}~\bibnamefont {Li}}, \bibinfo {author}
  {\bibfnamefont {B.}~\bibnamefont {Chen}}, \bibinfo {author} {\bibfnamefont
  {K.}~\bibnamefont {Watanabe}}, \bibinfo {author} {\bibfnamefont
  {T.}~\bibnamefont {Taniguchi}}, \bibinfo {author} {\bibfnamefont
  {S.}~\bibnamefont {Tongay}}, \bibinfo {author} {\bibfnamefont
  {G.}~\bibnamefont {Wang}}, \bibinfo {author} {\bibfnamefont {A.~C.}\
  \bibnamefont {Ferrari}}, \ and\ \bibinfo {author} {\bibfnamefont
  {M.}~\bibnamefont {Atat{\"u}re}},\ }\bibfield  {title} {\bibinfo {title}
  {\emph {Charge-tuneable biexciton complexes in monolayer WSe2}},\ }\href
  {\doibase 10.1038/s41467-018-05632-4} {\bibfield  {journal} {\bibinfo
  {journal} {Nature Communications}\ }\textbf {\bibinfo {volume} {9}},\
  \bibinfo {pages} {3721} (\bibinfo {year} {2018})}\BibitemShut {NoStop}%
\bibitem [{\citenamefont {Muir}\ \emph {et~al.}(2022)\citenamefont {Muir},
  \citenamefont {Levinsen}, \citenamefont {Earl}, \citenamefont {Conway},
  \citenamefont {Cole}, \citenamefont {Wurdack}, \citenamefont {Mishra},
  \citenamefont {Ing}, \citenamefont {Estrecho}, \citenamefont {Lu},
  \citenamefont {Efimkin}, \citenamefont {Tollerud}, \citenamefont
  {Ostrovskaya}, \citenamefont {Parish},\ and\ \citenamefont
  {Davis}}]{Muir2022}%
  \BibitemOpen
  \bibfield  {author} {\bibinfo {author} {\bibfnamefont {J.~B.}\ \bibnamefont
  {Muir}}, \bibinfo {author} {\bibfnamefont {J.}~\bibnamefont {Levinsen}},
  \bibinfo {author} {\bibfnamefont {S.~K.}\ \bibnamefont {Earl}}, \bibinfo
  {author} {\bibfnamefont {M.~A.}\ \bibnamefont {Conway}}, \bibinfo {author}
  {\bibfnamefont {J.~H.}\ \bibnamefont {Cole}}, \bibinfo {author}
  {\bibfnamefont {M.}~\bibnamefont {Wurdack}}, \bibinfo {author} {\bibfnamefont
  {R.}~\bibnamefont {Mishra}}, \bibinfo {author} {\bibfnamefont {D.~J.}\
  \bibnamefont {Ing}}, \bibinfo {author} {\bibfnamefont {E.}~\bibnamefont
  {Estrecho}}, \bibinfo {author} {\bibfnamefont {Y.}~\bibnamefont {Lu}},
  \bibinfo {author} {\bibfnamefont {D.~K.}\ \bibnamefont {Efimkin}}, \bibinfo
  {author} {\bibfnamefont {J.~O.}\ \bibnamefont {Tollerud}}, \bibinfo {author}
  {\bibfnamefont {E.~A.}\ \bibnamefont {Ostrovskaya}}, \bibinfo {author}
  {\bibfnamefont {M.~M.}\ \bibnamefont {Parish}}, \ and\ \bibinfo {author}
  {\bibfnamefont {J.~A.}\ \bibnamefont {Davis}},\ }\bibfield  {title} {\bibinfo
  {title} {\emph {Interactions between Fermi polarons in monolayer WS2}},\
  }\href {\doibase 10.1038/s41467-022-33811-x} {\bibfield  {journal} {\bibinfo
  {journal} {Nature Communications}\ }\textbf {\bibinfo {volume} {13}},\
  \bibinfo {pages} {6164} (\bibinfo {year} {2022})}\BibitemShut {NoStop}%
\bibitem [{\citenamefont {Yang}\ \emph
  {et~al.}(2015{\natexlab{b}})\citenamefont {Yang}, \citenamefont {Xu},
  \citenamefont {Pei}, \citenamefont {Myint}, \citenamefont {Wang},
  \citenamefont {Wang}, \citenamefont {Zhang}, \citenamefont {Yu},\ and\
  \citenamefont {Lu}}]{Yang2015phosphorene}%
  \BibitemOpen
  \bibfield  {author} {\bibinfo {author} {\bibfnamefont {J.}~\bibnamefont
  {Yang}}, \bibinfo {author} {\bibfnamefont {R.}~\bibnamefont {Xu}}, \bibinfo
  {author} {\bibfnamefont {J.}~\bibnamefont {Pei}}, \bibinfo {author}
  {\bibfnamefont {Y.~W.}\ \bibnamefont {Myint}}, \bibinfo {author}
  {\bibfnamefont {F.}~\bibnamefont {Wang}}, \bibinfo {author} {\bibfnamefont
  {Z.}~\bibnamefont {Wang}}, \bibinfo {author} {\bibfnamefont {S.}~\bibnamefont
  {Zhang}}, \bibinfo {author} {\bibfnamefont {Z.}~\bibnamefont {Yu}}, \ and\
  \bibinfo {author} {\bibfnamefont {Y.}~\bibnamefont {Lu}},\ }\bibfield
  {title} {\bibinfo {title} {\emph {Optical tuning of exciton and trion
  emissions in monolayer phosphorene}},\ }\href {\doibase 10.1038/lsa.2015.85}
  {\bibfield  {journal} {\bibinfo  {journal} {Light: Science \& Applications}\
  }\textbf {\bibinfo {volume} {4}},\ \bibinfo {pages} {e312} (\bibinfo {year}
  {2015}{\natexlab{b}})}\BibitemShut {NoStop}%
\bibitem [{\citenamefont {Xu}\ \emph {et~al.}(2016)\citenamefont {Xu},
  \citenamefont {Zhang}, \citenamefont {Wang}, \citenamefont {Yang},
  \citenamefont {Wang}, \citenamefont {Pei}, \citenamefont {Myint},
  \citenamefont {Xing}, \citenamefont {Yu}, \citenamefont {Fu}, \citenamefont
  {Qin},\ and\ \citenamefont {Lu}}]{Xu2016phosphorene}%
  \BibitemOpen
  \bibfield  {author} {\bibinfo {author} {\bibfnamefont {R.}~\bibnamefont
  {Xu}}, \bibinfo {author} {\bibfnamefont {S.}~\bibnamefont {Zhang}}, \bibinfo
  {author} {\bibfnamefont {F.}~\bibnamefont {Wang}}, \bibinfo {author}
  {\bibfnamefont {J.}~\bibnamefont {Yang}}, \bibinfo {author} {\bibfnamefont
  {Z.}~\bibnamefont {Wang}}, \bibinfo {author} {\bibfnamefont {J.}~\bibnamefont
  {Pei}}, \bibinfo {author} {\bibfnamefont {Y.~W.}\ \bibnamefont {Myint}},
  \bibinfo {author} {\bibfnamefont {B.}~\bibnamefont {Xing}}, \bibinfo {author}
  {\bibfnamefont {Z.}~\bibnamefont {Yu}}, \bibinfo {author} {\bibfnamefont
  {L.}~\bibnamefont {Fu}}, \bibinfo {author} {\bibfnamefont {Q.}~\bibnamefont
  {Qin}}, \ and\ \bibinfo {author} {\bibfnamefont {Y.}~\bibnamefont {Lu}},\
  }\bibfield  {title} {\bibinfo {title} {\emph {Extraordinarily Bound
  Quasi-One-Dimensional Trions in Two-Dimensional Phosphorene Atomic
  Semiconductors}},\ }\href {\doibase 10.1021/acsnano.5b06193} {\bibfield
  {journal} {\bibinfo  {journal} {ACS Nano}\ }\textbf {\bibinfo {volume}
  {10}},\ \bibinfo {pages} {2046} (\bibinfo {year} {2016})}\BibitemShut
  {NoStop}%
\bibitem [{\citenamefont {Li}\ \emph {et~al.}(2021)\citenamefont {Li},
  \citenamefont {Bleu}, \citenamefont {Parish},\ and\ \citenamefont
  {Levinsen}}]{Li2021PRL}%
  \BibitemOpen
  \bibfield  {author} {\bibinfo {author} {\bibfnamefont {G.}~\bibnamefont
  {Li}}, \bibinfo {author} {\bibfnamefont {O.}~\bibnamefont {Bleu}}, \bibinfo
  {author} {\bibfnamefont {M.~M.}\ \bibnamefont {Parish}}, \ and\ \bibinfo
  {author} {\bibfnamefont {J.}~\bibnamefont {Levinsen}},\ }\bibfield  {title}
  {\bibinfo {title} {\emph {Enhanced {Scattering} between {Electrons} and
  {Exciton}-{Polaritons} in a {Microcavity}}},\ }\href {\doibase
  10.1103/PhysRevLett.126.197401} {\bibfield  {journal} {\bibinfo  {journal}
  {Physical Review Letters}\ }\textbf {\bibinfo {volume} {126}},\ \bibinfo
  {pages} {197401} (\bibinfo {year} {2021})}\BibitemShut {NoStop}%
\bibitem [{\citenamefont {Kumar}\ \emph {et~al.}(2023)\citenamefont {Kumar},
  \citenamefont {Mulkerin}, \citenamefont {Parish},\ and\ \citenamefont
  {Levinsen}}]{Kumar2023}%
  \BibitemOpen
  \bibfield  {author} {\bibinfo {author} {\bibfnamefont {S.~S.}\ \bibnamefont
  {Kumar}}, \bibinfo {author} {\bibfnamefont {B.~C.}\ \bibnamefont {Mulkerin}},
  \bibinfo {author} {\bibfnamefont {M.~M.}\ \bibnamefont {Parish}}, \ and\
  \bibinfo {author} {\bibfnamefont {J.}~\bibnamefont {Levinsen}},\ }\bibfield
  {title} {\bibinfo {title} {\emph {Trion resonance in polariton-electron
  scattering}},\ }\href {\doibase 10.1103/PhysRevB.108.125416} {\bibfield
  {journal} {\bibinfo  {journal} {Phys. Rev. B}\ }\textbf {\bibinfo {volume}
  {108}},\ \bibinfo {pages} {125416} (\bibinfo {year} {2023})}\BibitemShut
  {NoStop}%
\bibitem [{\citenamefont {Plechinger}\ \emph {et~al.}(2016)\citenamefont
  {Plechinger}, \citenamefont {Nagler}, \citenamefont {Arora}, \citenamefont
  {Schmidt}, \citenamefont {Chernikov}, \citenamefont {del {\'A}guila},
  \citenamefont {Christianen}, \citenamefont {Bratschitsch}, \citenamefont
  {Sch{\"u}ller},\ and\ \citenamefont {Korn}}]{Plechinger2016}%
  \BibitemOpen
  \bibfield  {author} {\bibinfo {author} {\bibfnamefont {G.}~\bibnamefont
  {Plechinger}}, \bibinfo {author} {\bibfnamefont {P.}~\bibnamefont {Nagler}},
  \bibinfo {author} {\bibfnamefont {A.}~\bibnamefont {Arora}}, \bibinfo
  {author} {\bibfnamefont {R.}~\bibnamefont {Schmidt}}, \bibinfo {author}
  {\bibfnamefont {A.}~\bibnamefont {Chernikov}}, \bibinfo {author}
  {\bibfnamefont {A.~G.}\ \bibnamefont {del {\'A}guila}}, \bibinfo {author}
  {\bibfnamefont {P.~C.~M.}\ \bibnamefont {Christianen}}, \bibinfo {author}
  {\bibfnamefont {R.}~\bibnamefont {Bratschitsch}}, \bibinfo {author}
  {\bibfnamefont {C.}~\bibnamefont {Sch{\"u}ller}}, \ and\ \bibinfo {author}
  {\bibfnamefont {T.}~\bibnamefont {Korn}},\ }\bibfield  {title} {\bibinfo
  {title} {\emph {Trion fine structure and coupled spin--valley dynamics in
  monolayer tungsten disulfide}},\ }\href {\doibase 10.1038/ncomms12715}
  {\bibfield  {journal} {\bibinfo  {journal} {Nature Communications}\ }\textbf
  {\bibinfo {volume} {7}},\ \bibinfo {pages} {12715} (\bibinfo {year}
  {2016})}\BibitemShut {NoStop}%
\bibitem [{\citenamefont {Sidler}\ \emph {et~al.}(2017)\citenamefont {Sidler},
  \citenamefont {Back}, \citenamefont {Cotlet}, \citenamefont {Srivastava},
  \citenamefont {Fink}, \citenamefont {Kroner}, \citenamefont {Demler},\ and\
  \citenamefont {Imamoglu}}]{SidlerNatPhys16}%
  \BibitemOpen
  \bibfield  {author} {\bibinfo {author} {\bibfnamefont {M.}~\bibnamefont
  {Sidler}}, \bibinfo {author} {\bibfnamefont {P.}~\bibnamefont {Back}},
  \bibinfo {author} {\bibfnamefont {O.}~\bibnamefont {Cotlet}}, \bibinfo
  {author} {\bibfnamefont {A.}~\bibnamefont {Srivastava}}, \bibinfo {author}
  {\bibfnamefont {T.}~\bibnamefont {Fink}}, \bibinfo {author} {\bibfnamefont
  {M.}~\bibnamefont {Kroner}}, \bibinfo {author} {\bibfnamefont
  {E.}~\bibnamefont {Demler}}, \ and\ \bibinfo {author} {\bibfnamefont
  {A.}~\bibnamefont {Imamoglu}},\ }\bibfield  {title} {\bibinfo {title} {\emph
  {Fermi polaron-polaritons in charge-tunable atomically thin
  semiconductors}},\ }\href {https://doi.org/10.1038/nphys3949} {\bibfield
  {journal} {\bibinfo  {journal} {Nature Physics}\ }\textbf {\bibinfo {volume}
  {13}},\ \bibinfo {pages} {255} (\bibinfo {year} {2017})}\BibitemShut
  {NoStop}%
\bibitem [{\citenamefont {Efimkin}\ and\ \citenamefont
  {MacDonald}(2017)}]{Efimkin2017}%
  \BibitemOpen
  \bibfield  {author} {\bibinfo {author} {\bibfnamefont {D.~K.}\ \bibnamefont
  {Efimkin}}\ and\ \bibinfo {author} {\bibfnamefont {A.~H.}\ \bibnamefont
  {MacDonald}},\ }\bibfield  {title} {\bibinfo {title} {\emph {{Many-body
  theory of trion absorption features in two-dimensional semiconductors}}},\
  }\href {\doibase 10.1103/PhysRevB.95.035417} {\bibfield  {journal} {\bibinfo
  {journal} {Phys. Rev. B}\ }\textbf {\bibinfo {volume} {95}},\ \bibinfo
  {pages} {1} (\bibinfo {year} {2017})}\BibitemShut {NoStop}%
\bibitem [{\citenamefont {Huang}\ \emph {et~al.}(2023)\citenamefont {Huang},
  \citenamefont {Sampson}, \citenamefont {Ni}, \citenamefont {Liu},
  \citenamefont {Liang}, \citenamefont {Watanabe}, \citenamefont {Taniguchi},
  \citenamefont {Li}, \citenamefont {Martin}, \citenamefont {Levinsen},
  \citenamefont {Parish}, \citenamefont {Tutuc}, \citenamefont {Efimkin},\ and\
  \citenamefont {Li}}]{Huang2023}%
  \BibitemOpen
  \bibfield  {author} {\bibinfo {author} {\bibfnamefont {D.}~\bibnamefont
  {Huang}}, \bibinfo {author} {\bibfnamefont {K.}~\bibnamefont {Sampson}},
  \bibinfo {author} {\bibfnamefont {Y.}~\bibnamefont {Ni}}, \bibinfo {author}
  {\bibfnamefont {Z.}~\bibnamefont {Liu}}, \bibinfo {author} {\bibfnamefont
  {D.}~\bibnamefont {Liang}}, \bibinfo {author} {\bibfnamefont
  {K.}~\bibnamefont {Watanabe}}, \bibinfo {author} {\bibfnamefont
  {T.}~\bibnamefont {Taniguchi}}, \bibinfo {author} {\bibfnamefont
  {H.}~\bibnamefont {Li}}, \bibinfo {author} {\bibfnamefont {E.}~\bibnamefont
  {Martin}}, \bibinfo {author} {\bibfnamefont {J.}~\bibnamefont {Levinsen}},
  \bibinfo {author} {\bibfnamefont {M.~M.}\ \bibnamefont {Parish}}, \bibinfo
  {author} {\bibfnamefont {E.}~\bibnamefont {Tutuc}}, \bibinfo {author}
  {\bibfnamefont {D.~K.}\ \bibnamefont {Efimkin}}, \ and\ \bibinfo {author}
  {\bibfnamefont {X.}~\bibnamefont {Li}},\ }\bibfield  {title} {\bibinfo
  {title} {\emph {Quantum Dynamics of Attractive and Repulsive Polarons in a
  Doped ${\mathrm{MoSe}}_{2}$ Monolayer}},\ }\href {\doibase
  10.1103/PhysRevX.13.011029} {\bibfield  {journal} {\bibinfo  {journal} {Phys.
  Rev. X}\ }\textbf {\bibinfo {volume} {13}},\ \bibinfo {pages} {011029}
  (\bibinfo {year} {2023})}\BibitemShut {NoStop}%
\bibitem [{\citenamefont {Jones}\ \emph {et~al.}(2013)\citenamefont {Jones},
  \citenamefont {Yu}, \citenamefont {Ghimire}, \citenamefont {Wu},
  \citenamefont {Aivazian}, \citenamefont {Ross}, \citenamefont {Zhao},
  \citenamefont {Yan}, \citenamefont {Mandrus}, \citenamefont {Xiao},
  \citenamefont {Yao},\ and\ \citenamefont {Xu}}]{Jones2013}%
  \BibitemOpen
  \bibfield  {author} {\bibinfo {author} {\bibfnamefont {A.~M.}\ \bibnamefont
  {Jones}}, \bibinfo {author} {\bibfnamefont {H.}~\bibnamefont {Yu}}, \bibinfo
  {author} {\bibfnamefont {N.~J.}\ \bibnamefont {Ghimire}}, \bibinfo {author}
  {\bibfnamefont {S.}~\bibnamefont {Wu}}, \bibinfo {author} {\bibfnamefont
  {G.}~\bibnamefont {Aivazian}}, \bibinfo {author} {\bibfnamefont {J.~S.}\
  \bibnamefont {Ross}}, \bibinfo {author} {\bibfnamefont {B.}~\bibnamefont
  {Zhao}}, \bibinfo {author} {\bibfnamefont {J.}~\bibnamefont {Yan}}, \bibinfo
  {author} {\bibfnamefont {D.~G.}\ \bibnamefont {Mandrus}}, \bibinfo {author}
  {\bibfnamefont {D.}~\bibnamefont {Xiao}}, \bibinfo {author} {\bibfnamefont
  {W.}~\bibnamefont {Yao}}, \ and\ \bibinfo {author} {\bibfnamefont
  {X.}~\bibnamefont {Xu}},\ }\bibfield  {title} {\bibinfo {title} {\emph
  {Optical generation of excitonic valley coherence in monolayer {WSe2}}},\
  }\href {\doibase 10.1038/nnano.2013.151} {\bibfield  {journal} {\bibinfo
  {journal} {Nature Nanotechnology}\ }\textbf {\bibinfo {volume} {8}},\
  \bibinfo {pages} {634} (\bibinfo {year} {2013})}\BibitemShut {NoStop}%
\bibitem [{\citenamefont {Li}\ \emph {et~al.}(2022)\citenamefont {Li},
  \citenamefont {Goryca}, \citenamefont {Choi}, \citenamefont {Xu},\ and\
  \citenamefont {Crooker}}]{Li2021Exciton}%
  \BibitemOpen
  \bibfield  {author} {\bibinfo {author} {\bibfnamefont {J.}~\bibnamefont
  {Li}}, \bibinfo {author} {\bibfnamefont {M.}~\bibnamefont {Goryca}}, \bibinfo
  {author} {\bibfnamefont {J.}~\bibnamefont {Choi}}, \bibinfo {author}
  {\bibfnamefont {X.}~\bibnamefont {Xu}}, \ and\ \bibinfo {author}
  {\bibfnamefont {S.~A.}\ \bibnamefont {Crooker}},\ }\bibfield  {title}
  {\bibinfo {title} {\emph {Many-Body Exciton and Intervalley Correlations in
  Heavily Electron-Doped WSe2 Monolayers}},\ }\href {\doibase
  10.1021/acs.nanolett.1c04217} {\bibfield  {journal} {\bibinfo  {journal}
  {Nano Letters}\ }\textbf {\bibinfo {volume} {22}},\ \bibinfo {pages} {426}
  (\bibinfo {year} {2022})}\BibitemShut {NoStop}%
\end{thebibliography}%

\end{document}